\documentclass[pre, superscriptaddress, longbibliography, amsmath, amssymb, aps, twocolumn]{revtex4-2}
\usepackage{xr}
\usepackage{graphicx}
\usepackage{dcolumn}
\usepackage{bm}
\usepackage{color}
\usepackage{multibib}

\makeatletter

\begin{document}

\title{Generating active T1 transitions through mechanochemical feedback}

\author{Rastko Sknepnek}
\email{r.sknepnek@dundee.ac.uk}
\affiliation{School of Science and Engineering, University of Dundee, Dundee DD1 4HN, United Kingdom}
\affiliation{School of Life Sciences, University of Dundee, Dundee DD1 5EH, United Kingdom}
\author{Ilyas Djafer-Cherif}
\affiliation{School of Mathematics, University of Bristol, Bristol BS8 1UG, United Kingdom}
\affiliation{Dioscuri Centre for Physics and Chemistry of Bacteria, Institute of Physical Chemistry, Polish Academy of Sciences, 01-224 Warsaw, Poland}
\author{Manli Chuai}
\affiliation{School of Life Sciences, University of Dundee, Dundee DD1 5EH, United Kingdom}
\author{Cornelis J.~Weijer}
\email{c.j.weijer@dundee.ac.uk}
\affiliation{School of Life Sciences, University of Dundee, Dundee DD1 5EH, United Kingdom}
\author{Silke Henkes}
\email{shenkes@lorentz.leidenuniv.nl}
\affiliation{School of Mathematics, University of Bristol, Bristol BS8 1UG, United Kingdom}
\affiliation{Lorentz Institute, Leiden University, 2300RA Leiden, The Netherlands}

\begin{abstract}
    Convergence-extension in embryos is controlled by chemical and mechanical signalling. A key cellular process is the exchange of neighbours via T1 transitions. We propose and analyse a model with positive feedback between recruitment of myosin motors and mechanical tension in cell junctions. The model produces active T1 events, which act to elongate the tissue perpendicular to external pulling. Using an idealized tissue patch comprising several active cells embedded in a matrix of passive hexagonal cells we identified an optimal range of pulling forces to trigger an active T1 event. We show that pulling also generates tension chains in a realistic patch made entirely of active cells of random shapes, and leads to convergence-extension over a range of parameters. Our findings show that active intercalations can generate stress that activates T1 events in neighbouring cells resulting in tension dependent tissue re-organisation, in qualitative agreement with experiments on gastrulation in chick embryos.
\end{abstract}

\maketitle

\section{Introduction}

Embryonic development involves complex tissue dynamics, including rearrangements and shape changes of the cells. This is particularly evident during gastrulation where the presumptive ectoderm, mesoderm, and endoderm take up their correct positions in the embryo \cite{wolpert2015principles}. Key cellular processes that underlie tissue formation and morphogenesis during gastrulation are cell division, differentiation, and cell movement. Directed cell intercalation, i.e. polarised movement of cells past each other, is a major mechanism driving large scale tissue shape changes both in epithelial and mesenchymal tissues \cite{huebner2018coming}. The narrowing and lengthening of epithelial tissues resulting from such intercalations, known as convergent extension \cite{keller2000mechanisms}, underlies germband extension in Drosophila as well as primitive streak formation in the chick embryo \cite{voiculescu2007amniote,rozbicki2015myosin}. In the latter, cell intercalations facilitate coordinated movements of hundreds of thousands of cells in two counter-rotating millimetre scale cell flows that drive the formation of the primitive streak at the site where the flows meet \cite{rozbicki2015myosin,saadaoui2020tensile,najera2020cellular}. Unlike cell migration \cite{alert2020physical}, which typically involves a significant contribution from crawling against a substrate such as the extracellular matrix, during intercalation, cells pull against each other in order to exchange their neighbours \cite{huebner2018coming}. This is a complex, active process that requires a carefully coordinated shrinking and subsequent expansion of cell-cell interfaces, known as the T1 transition \cite{fernandez2009myosin}.

With advances in in-vivo imaging of early stage chick embryos \cite{tomer2012quantitative,chen2014lattice,rozbicki2015myosin}, it is possible to track behaviours of individual cells. These experiments show that the proper formation of large-scale structures intricately depends on coordination of intercalation, ingression, and division events \cite{firmino2016cell,najera2020cellular}. Although the specific details are species-dependent, the underlying machinery is evolutionarily conserved and in its core relies on force generation provided by the cell's actomyosin cortex \cite{lecuit2011force}. The cell-level mechanisms behind those coordinated events are, however, poorly understood. With cellular behaviours being coordinated over thousands of cells in the case of the chick embryo, biochemical signalling alone is unlikely to account for the observed motion patterns. 

Experiments on Drosophila germband extension have shown that cell intercalations are a result of a combined action of junctional and medial myosin \cite{bertet2004myosin,blankenship2006multicellular,rauzi2010planar}. Junctional myosin is localised at the apical junctional complex, a structure that encircles the cell and tightly links it through cell-cell adhesion molecules to its neighbours. It generates tension along the junction that can act to shrink it \cite{rauzi2008nature,fernandez2009myosin}. Furthermore, junctional myosin in neighbouring cells has been shown to form cable-like supercellular structures \cite{jacinto2002dynamic,fernandez2009myosin,rauzi2010planar,legoff2013global,sugimura2013mechanical,clement2017viscoelastic} that have been shown to correspond to increased tension \cite{fernandez2009myosin}.

\begin{figure*}[htb]
    \centering
    \includegraphics[width=0.95\textwidth]{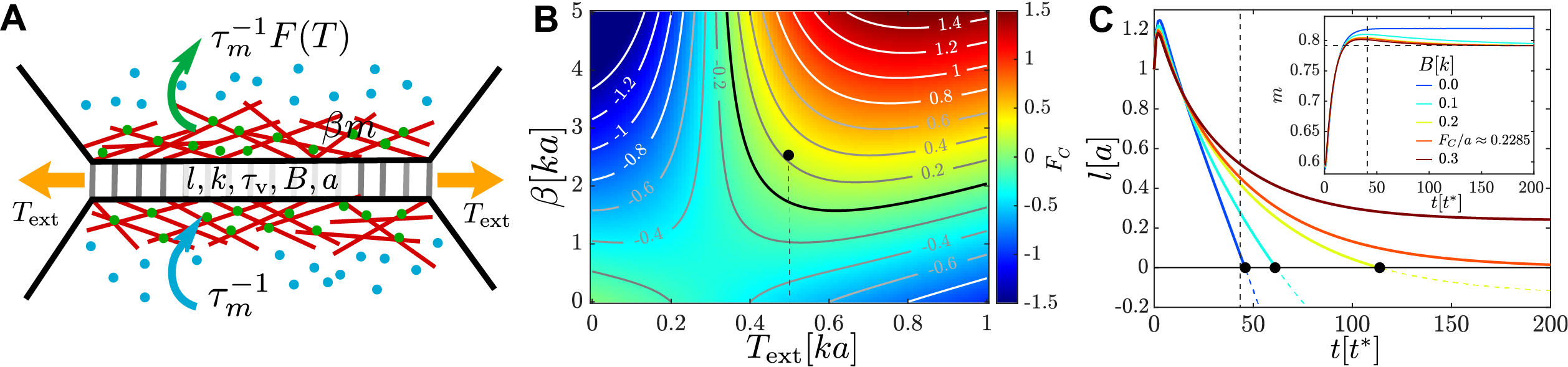}
    \caption{\textbf{A}) An external pulling force of magnitude $T_\text{ext}$ induces tension $T$ in a cell-cell junction of length $l$, which consists of passive viscoelastic and active components. The passive component consists of a Maxwell element with stiffness $k$ and viscous relaxation time $\tau_\text{v}$ and a harmonic spring of stiffness $B$ and rest length $a$ connected to it in parallel. The active component is due to myosin motors (green and blue dots) with concentration $m$ that act to contract cortical actin filaments (red lines), exerting a force of magnitude $\beta m$. Myosin motors bind to the actin cortex with association rate $\tau_\text{m}^{-1}$ and unbind with a tension-dependent dissociation rate $\tau_\text{m}^{-1} F(T)$. \textbf{B}) Heatmap plot of the contraction force $F_\text{C}\left(T_\text{ext},\beta\right)$. For $B=0$, the junction contraction rate is $\dot{l} = F_\text{C}/\zeta$, where $\zeta$ is the friction coefficient with the surrounding medium. The mechanochemical feedback loop is contractile in the top right quadrant where $\beta>\beta_\text{c}$, $T_\text{ext}>T^*$, and $F_\text{C}>0$. Negative values of $F_\text{C}$ correspond to an extending junction. \textbf{C}) Junction length vs.~time for $T_\text{ext}=0.5ka$, $\beta=2.5ka$, $\tau_\text{v}=\tau_\text{m}=10t^*$ (black dot in B) for increasing values of the elastic barrier $B$. An active $T1$ corresponds to reaching $l=0$. Increasing $B$ slows down contractions, until, for $B \geq F_\text{C}/a$, the equilibrium length $l\ge0$ and no T1 is possible. Inset: Myosin dynamics for the same set of junctions; the horizontal dashed line indicates $m_\text{eq}$. $\alpha=1$, $T^*=0.3ka$, $k_0=2/T^*$, and $m_0=0.5$. Length is measured in units of $a$, time in units of $t^*=\zeta/k$, and force in units of $ka$. }
    \label{fig:junction_barrier}
\end{figure*}

To develop a cell-level model of the convergence-extension process, it is necessary to understand how externally applied and internally generated mechanical stresses couple to the signalling pathways that regulate the cell's mechanical response. One, therefore, needs to understand the feedback between mechanical stress anisotropy and polarity in the distribution of force-generating molecular motors in the cell, i.e.~how it emerges and is propagated and coordinated over large distances. The observation that coupling between chemical processes and mechanical responses plays an important role during embryonic development dates back more than 40 years \cite{beloussov1975mechanical,odell1980mechanical,belintsev1987model}, but has also been discussed more recently \cite{bailles2019genetic}. While several models that address different aspects of the mechanochemical coupling have been proposed \cite{zajac2003simulating,wang2012cell,spahn2013vertex,lan2015biomechanical,shindo2019pcp,zankoc2020elasticity}, many of them assume the existence of a chemical prepattern that drives the initial symmetry breaking, e.g.~by explicitly enhancing tension in cell-cell junctions in a preferred direction. The initial chemical polarity is indeed present in some organisms, however, there are systems, e.g.~early-stage chick embryos, where no such chemical prepatterning controlling tension has been found, and the symmetry breaking is driven by a different mechanism such as mechanical polarity. The origin of the mechanical polarisation is not fully understood. 

The aim of this study is to formulate and analyse a model for cell intercalations that includes explicit mechanochemical coupling and does not require initial chemical prepatterning. The initial symmetry breaking is driven by mechanical rather than chemical polarisation. The focus is on the active T1 transition, which occurs perpendicular to the direction of the maximum principal mechanical stress. In the model, this stress is assumed to be anisotropic and externally applied, while in an embryo it is produced by the tissue surrounding the region of interest, e.g. the sickle-shaped region in the posterior of the chick embryo that develops into the primitive streak \cite{najera2020cellular}. Unlike passive T1 events that are local plastic rearrangements that relieve the applied stresses as, e.g. in foams \cite{weaire2001physics}, active T1 transitions require the cell to induce junction contractions via self-amplifying generation of tension. The key ingredient of the model is, therefore, a feedback mechanism between the kinetics of the force producing molecules, here assumed to be myosin, and mechanical tension in cell junctions. 

Here we construct a model that generically provides a mechanism for active T1 events that underlie convergent extension flows such as those observed during primitive streak formation in the chick embryo \cite{rozbicki2015myosin}. Our analysis indicates that the viscoelastic nature of the cell-cell junctions is essential for an active T1 event, in agreement with studies on ratcheting during junction contractions \cite{clement2017viscoelastic, staddon2019mechanosensitive}. In addition, for the active T1 transition to be possible, there must be a separation of elastic ($t^*$), viscoelastic remodelling ($\tau_\text{v}$), and motor turnover timescales ($\tau_\text{m}$), with $\tau_\text{v},\tau_\text{m} > t^*$. 

We first analyse the mechanochemical feedback in the case of a single junction, which describes the key ingredients of the proposed mechanism but avoids complications associated with cell rearrangements. The analysis then proceeds to the two-dimensional case, implemented as an extension of the vertex model~\cite{farhadifar2007influence,fletcher2014vertex}, where cell rearrangement are not only possible, but lead to shape changes at the tissue scale. We find active T1 transitions and convergence-extension flows over a broad region of externally applied stresses and relaxation time scales, confirming that the proposed mechanism is robust.

\section*{Results}

{\bf Single-junction model.} To understand the mechanism that couples the kinetics of myosin motors to the local mechanical tension and leads to the activation of contractility in cell-cell junctions, we first analyse a model of a single junction. The single-junction model thus provides insight into the conditions under which the junction length can contract to zero and trigger a T1 transition. In this model, for simplicity, the junction is assumed to be surrounded by a tissue which provides an elastic, tension-generating background against which it actively contracts. Guided by experiments, there are three key ingredients of the single-junction model. 1) The junction is viscoelastic, as established by pull-release optical tweezer experiments on cell-cell junctions in Drosophila and chick embryos~\cite{clement2017viscoelastic,ferro2020measurement}. This means that the junction is able to remove imposed tension by remodelling itself. 2) The junction can generate tension via the action of myosin motor minifilaments that slide actin filaments against each other. 3) There is self-amplifying feedback due to the exponentially decreasing unbinding rate of myosin motors with tension in the junction~\cite{veigel2003load,kovacs2007load}. A graphic representation of the model is shown in fig.~\ref{fig:junction_barrier}A and details are discussed in Sec.~\ref{subsec:SI-single-junction} in Supplementary Information, with parameter values given in tab.~\ref{tab:single-junction}. 

\begin{figure*}[tb]
    \centering
    \includegraphics[width=0.99\textwidth]{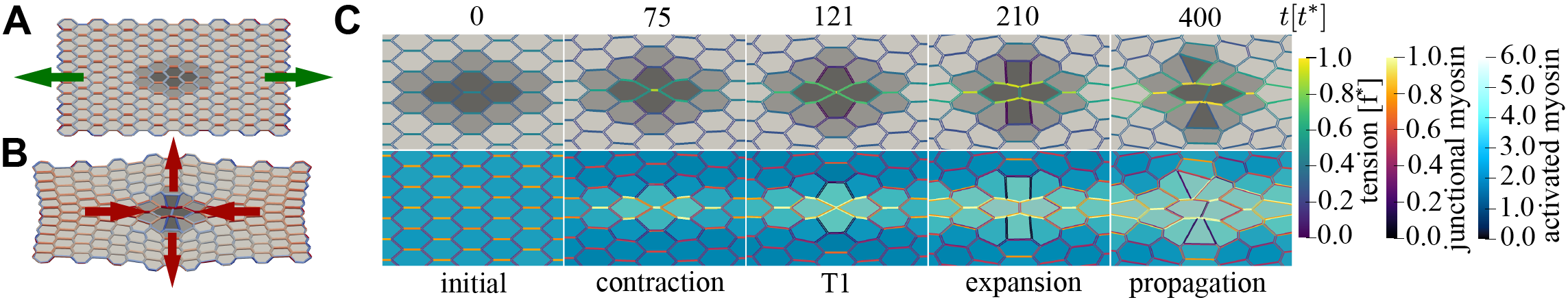}
    \caption{An active T1 transition event. \textbf{A}) The initial state is mechanically polarised by applying pulling forces (green arrows) in the horizontal direction to the left and right boundaries. \textbf{B}) The final state after the active T1 shows a clear convergence-extension deformation (red arrows). Cells are coloured by type: passive (light grey), buffer (medium grey), and active (dark grey). Junctions are coloured by junctional myosin. \textbf{C}) Time sequence of the active T1 transition measured from the moment activity and viscoelasticity were switched on. Cells in the top row are coloured by type and junctions are coloured by tension. Cells in the bottom row are coloured by activated myosin $m_{\text{act}}^C$, and junctions are coloured by myosin. Parameters: $A_0 = \frac{3}{2}\sqrt{3}a^2$, $P_0=6a$, $\beta = 0.8f^*$ (active), $\beta = 0.4f^*$ (buffer), $\beta = 0$ (passive), $M=6$, $T^* = 0.3f^*$, $k_0 = 2/T^*$, $\tau_v = 20t^*$, $\tau_m = 100t^*$, $\alpha=0.1$, $f=1$, $f_\text{pull} = 0.15f^*$, with $n_x=15$ ($n_y=11$) cells in the horizontal (vertical) direction. Units: length ($a$), time ($t^*=\zeta/\left(\Gamma+k\right)$), force ($f^*=\left(\Gamma+k\right)a$). }
    \label{fig:T1_schematic}
\end{figure*}

We adopt one of the simplest descriptions that includes these three ingredients and assume that the junction consists of four elements connected in parallel: 1) A Maxwell element with stiffness $k$ and viscous relaxation timescale $\tau_\text{v}$, which models the viscoelastic character of the junction; 2) An elastic spring with spring constant $B$ and rest length $a$, which represents the elastic background, i.e. the \emph{elastic barrier}; 3) An active element that models the contribution of the cytoskeleton by generating active tension $\beta\left(m-m_0\right)$, where $\beta$ is the activity, $m$ is the ratio of the number of myosin motors bound to the junction and the maximum possible number of bound motors, and $m_0$ is the reference value of $m$; and 4) A dashpot with dissipation rate $1/\zeta$, which models dissipation with the surrounding medium. The first two elements form a Standard Linear Solid (SLS) element (see fig.~\ref{fig:SI_model}A). The presence of $m_0$ in the active element is necessary to account for the possibility that active contractions of the surrounding tissue are stronger than those in the junction, which would result in it expanding. Furthermore, since $m$ and $m_0$ measure the relative numbers of bound motors, the expression for the active force does not include the junction length (for further discussion, see Sec.~\ref{subsec:SI-single-junction} in the Supplementary Information). Under these assumptions, the dynamics of a junction with length $l$ and the rest length $l_0$ is
\begin{equation}
 \zeta \dot{l} = -T + T_{\text{ext}},  \quad \tau_\text{v} \dot{l}_0 = l-l_0, \quad \tau_\text{m} \dot{m} =  1 - m F(T),  \label{eq:one_junction} 
\end{equation}
where $T = k \left(l-l_0\right) + \beta\left(m-m_0\right) + B\left(l-a\right)$ is the junction tension, and $T_\text{ext}$ is external tension. The feedback loop between the concentration of bound myosin motors and the mechanical tension is captured by the equation for the myosin dynamics, which incorporates a tension-independent myosin binding rate $\tau_m^{-1}$, and an unbinding rate $F(T)/\tau_m$ that decreases with tension as $F(T) = \alpha + e^{-k_0(T-T^*)}$.
At steady state $\dot{m}=0$, and the equilibrium myosin $m_\text{eq}=F(T)^{-1}$ is a sigmoid function of tension. $T^*$, therefore, sets the threshold that separates low and high levels of attached myosin motors. $\alpha$ and $k_0$ are constants with choices of their values discussed in Sec.~\ref{subsec:SI-single-junction} in the Supplementary Information.

The first two equations in (\ref{eq:one_junction}) can be combined as $\zeta \dot{u}  = -\frac{\zeta}{\tau_\text{v}} u - T + T_{\text{ext}}$, where $u=l-l_0$. The intersection of nullclines $\dot{u}=0$ and $\dot{m}=0$ defines the fixed points of the dynamics, $\left(m_{\text{eq}},u_{\text{eq}}\right)$ (see fig.~S1C). Experiments of Clement, et al.~\cite{clement2017viscoelastic}, showed that prolonged pulling forced the junction to remodel and retain the elongated shape. Therefore, the relevant regime consistent with observations in real tissues is where the viscoelastic remodelling and myosin association time scales are longer than the elastic relaxation timescale, i.e. for $\tau_\text{v},\tau_\text{m}>\zeta/k\equiv t^*$. For $B=0$, there is a unique stable fixed point $G0$ that determines the long-time dynamics of the junction and the length of the junction continues to change at a constant rate $\dot{l} = \dot{u} + \dot{l}_0 = u_{eq}/\tau_\text{v}$ (fig.~\ref{fig:junction_barrier}C, the $B=0$ curve). For a junction with no external load, therefore, a fixed point with $u_\text{eq}\le0$ corresponds to steady contraction, while a fixed point with $u_\text{eq}>0$ corresponds to expansion. 

In the presence of a finite elastic barrier of height $Ba$, the junction behaves as an elastic solid in the long-time limit, and it is in mechanical equilibrium with $T=T_{\text{ext}}$. This corresponds to a single steady state solution of eqns.~(\ref{eq:one_junction}) at $u_\text{eq}^B=0$, indicated by the fixed point GB in fig.~\ref{fig:SI_model}C. The corresponding steady state value of myosin, 
\begin{equation} 
    m_\text{eq} = \frac{1}{\alpha+e^{k_0 \left(T^*-T_{\text{ext}}\right)}}, \label{eq:m_eq}
\end{equation}
is independent of $B$, reflecting the fact that in mechanical equilibrium external tension is balanced by the tension generated by myosin motors. The condition for a T1 transition to occur is, therefore, that the active tension due to myosin motors is sufficiently strong to shrink the junction to $l_0=0$. The mechanical equilibrium condition $T=T_{\text{ext}}$ at that point allows one to compute the magnitude of the contraction force $F_\text{C}$ the junction generates, or equivalently, the maximum barrier height, $F_\text{C}\left(T_\text{ext},\beta\right) \equiv Ba  = \beta\left(m_\text{eq} - m_0\right) - T_\text{ext}$ that a contracting junction can overcome. Figure~\ref{fig:junction_barrier}B shows isolines of  $F_\text{C}$, where positive values of $F_\text{C}$ correspond to junctions that can contract down to a T1 in the presence of a load, while negative values of $F_\text{C}$ correspond to junctions that cannot. Above a threshold in $T_{\text{ext}}\gtrsim T^*$ and $\beta>\beta_\text{C}$, the junction is able to gradually generate sufficiently large contraction forces required to overcome the elastic barrier and shrink down. Conversely, for $T_{ext} \lesssim T^*$ the junction expands, which is the appropriate regime for elongation after a T1. Therefore, there is positive feedback between mechanical tension and activity, which results from the assumption that the myosin association rate is independent of tension while the dissociation rate decays exponentially with it. The isoline $F_\text{C}=0$, marked as a thick black curve in fig.~\ref{fig:junction_barrier}B, separates the contracting and the expanding regimes, and it corresponds to a critical threshold $\beta_\text{c}$ for a T1 transition, 
\begin{equation} 
    \beta_\text{c} \ge \frac{T_{\text{ext}}}{m_\text{eq}-m_0}, \label{eq:beta_c}
\end{equation}
where $m_\text{eq}$ is given by eqn.~(\ref{eq:m_eq}). Figure \ref{fig:junction_barrier}C shows the junction dynamics in the presence of barriers of different heights for $T_\text{ext}=0.5ka$, $\beta=2.5ka$ with $F_\text{C}\approx0.2285ka$, indicated by the black dot in fig.~\ref{fig:junction_barrier}B. The junction shrinks to a point for $B \leq F_\text{C}/a$, while for larger values of $B$ the contraction stops at finite $l$. The initial elongation of the junction in fig.~\ref{fig:junction_barrier}C is due to our choice of the initial value of $m$ and reflects the fact that it takes $\approx\tau_\text{m}$ for the active contractile machinery to kick in. 

We conclude by emphasising the contractile response to applied tension of the single-junction model that will be at the heart of convergence-extension mechanism discussed below. That is, applying small external forces will lead to an expanding junction. Increasing the force, however, leads to the junction shrinking due to the increase of bound myosin motors. Additionally, the contraction rate strongly depends on the timescale of viscoelastic relaxation in the junction. Both effects should be measurable experimentally.

\begin{figure*}[tb]
    \centering
    \includegraphics[width=0.95\textwidth]{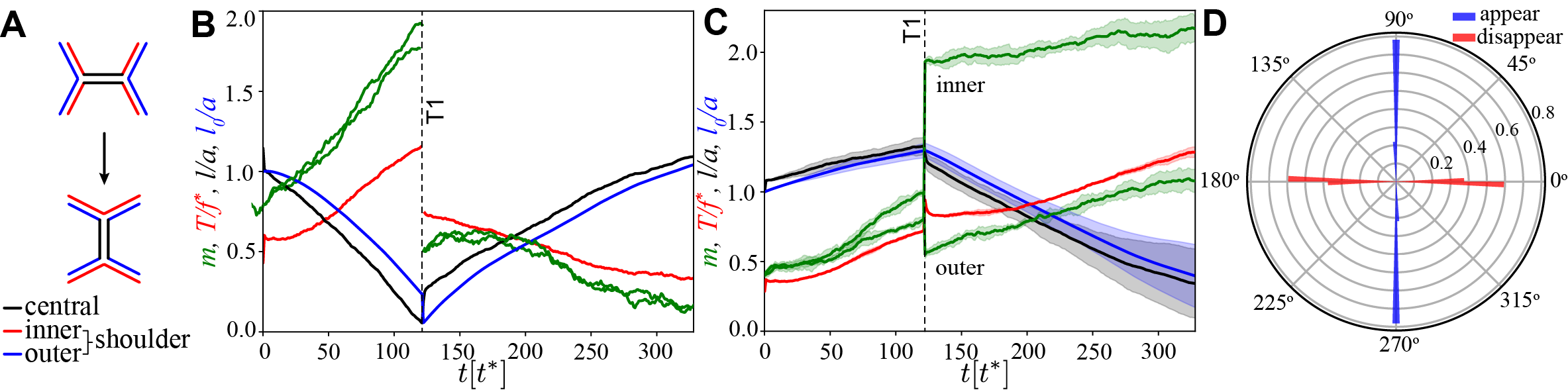}
    \caption{Junction dynamics during the active T1 transition shown in fig.~\ref{fig:T1_schematic}. \textbf{A}) Definition of central, inner and outer shoulder junctions through the T1 transition. \textbf{B}) Central junction: myosin, $m$ (green; two curves for myosin on two sides of the junction), tension, $T$ (red), junction length, $l$ (black), and rest length $l_0$ (blue) vs.~time. The vertical line indicates the T1 transition, at which point junctional myosin is redistributed according to the rules outlined in fig.~\ref{fig:AJM_schematic}B in Supplementary Information. \textbf{C}) Same as in panel A but averaged over four shoulder junctions, with variance indicated as shade. \textbf{D}) Polar histogram of the orientation of the first T1 event measured with respect to the pulling direction, from $N=32$ simulations. Blue (red) indicates appearing (disappearing) junctions. Parameters are $\beta=0.8f^*$ and $f_\text{pull}=0.15f^*$ and as in fig. \ref{fig:T1_schematic}. }
    \label{fig:T1_dynamics}
\end{figure*}

{\bf Vertex Model with Active Junctions.} The single junction model serves as a building block for a model of the entire epithelial tissue. A natural way to proceed is to extend an existing model for tissue mechanics. Setting aside apicobasal polarity, which affects cell intercalations in real tissues~\cite{huebner2018coming}, it can be assumed that the mechanical properties of the epithelial tissue arise chiefly from the apical junction cortex, an approximation that is able to qualitatively capture many aspects of tissue mechanics~\cite{fletcher2014vertex,murisic2015discrete}. 

We, therefore, model the mechanical response of the tissue with the vertex model~\cite{farhadifar2007influence,fletcher2014vertex}. The appeal of the vertex model is that it is able to capture both fluid and solid behaviours of epithelial tissues, including some aspects of the viscoelastic rheology \cite{tong2021linear, tong2022normal}. In the vertex model, the transition between the fluid state where cells can easily intercalate and tissue-scale flows are possible, and the solid phase where intercalations are arrested is controlled by a single dimensionless geometric parameter, $p_0 = P_0/\sqrt{A_0}$ \cite{staple2010mechanics,bi2015density,park2015unjamming,bi2016motility}. 

Here we extend the vertex model to include activity via the mechanochemical coupling introduced for the single junction and define the vertex model with active junctions. Details of the model are given in Sec.~\ref{sec:SI-Vertex-Model-with-Active-Junctions} in Supplementary Information. Each junction, shared by two cells denoted as $1$ and $2$, is augmented by two active elements that model active contractions by the actomyosin cortex on either side. The tension in the junction is then 
\begin{equation}
 T = T^{\text{P}} + k \left(l - l_0\right) + \beta_1 \left(m_1-m_0\right) + \beta_2 \left(m_2-m_0\right) , \label{eq:active_tension_main}
\end{equation}
where $T^\text{P}$ is the passive contribution from the standard vertex model energy given in eqn.~(\ref{eq:motion_VM}) in Supplementary Information. The junction rest length $l_0$ is, however, not constant but subject to viscoelastic relaxation with $\tau_\text{v} \dot{l}_0 = l-l_0$. It is important to note that $T^\text{P}$ depends only on the cell perimeter and not on junction length $l$ and, therefore does not allow for mechanical polarisation in response to an applied anisotropic tension. The additional spring term of the Maxwell element in eqn.~(\ref{eq:active_tension_main}) is, therefore, crucial for generating tension polarisation. The myosin dynamics of each active element is coupled to a conserved myosin pool of size $M$ for each cell $C$ that determines the association rate of myosin motors to the junctions. As in the single-junction model, the myosin dissociation rate is modulated by mechanical tension, leading to the following equation for myosin kinetics
\begin{equation}
    \tau_\text{m}\dot{m} = \left(M-m_\text{act}^\text{C}\right) - zm F\left(T\right) + \eta. \label{eq:myodyn_main}
\end{equation}
Here $m_\text{act}^\text{C} = \sum_{e=1}^z m_e$ is the total amount of activated myosin bound to the $z$ junctions of cell $C$, and we have included a noise component $\eta$ with zero mean and variance $f$ to model stochastic binding and unbinding of myosin. The overdamped dynamics of vertices is determined by force balance between friction, elastic forces due to deformations of the passive vertex model, and active forces due to pairs of active elements acting along the junctions connected to the vertex, i.e. $\zeta \dot{\mathbf{r}}_\text{i} = -\nabla_{\mathbf{r}_\text{i}}E_\text{VM} + \mathbf{F}^{\text{act}}_\text{i}$, where $\mathbf{r}_\text{i}$ is the position of vertex $i$, $E_\text{VM}$ is the energy of the passive vertex model (eqn. (\ref{eq:VM}) in Supplementary Information), and the active term $\mathbf{F}^{\text{act}}_i$ derives from the active elements introduced in eqn.~(\ref{eq:active_tension_main}).

{\bf Single active T1 transition in a hexagonal patch.} We begin by discussing a single active T1 event, as summarised in figs.~\ref{fig:T1_schematic} and \ref{fig:T1_dynamics}. The unit of time is set by the elastic timescale $t^* = \zeta/\left(k+\Gamma\right)$, the unit of length by the side a regular hexagon $a$, and the unit of force by $f^*=\left(\Gamma+k\right)a$, where $\Gamma$ is the perimeter modulus of the passive vertex model introduced in eqn. (\ref{eq:VM}) in Supplementary Information. Values of the parameters used in simulations are listed in tab.~\ref{tab:ajm} in Supplementary Information.

\begin{figure*}[tb]
    \centering
    \includegraphics[width=1.0\textwidth]{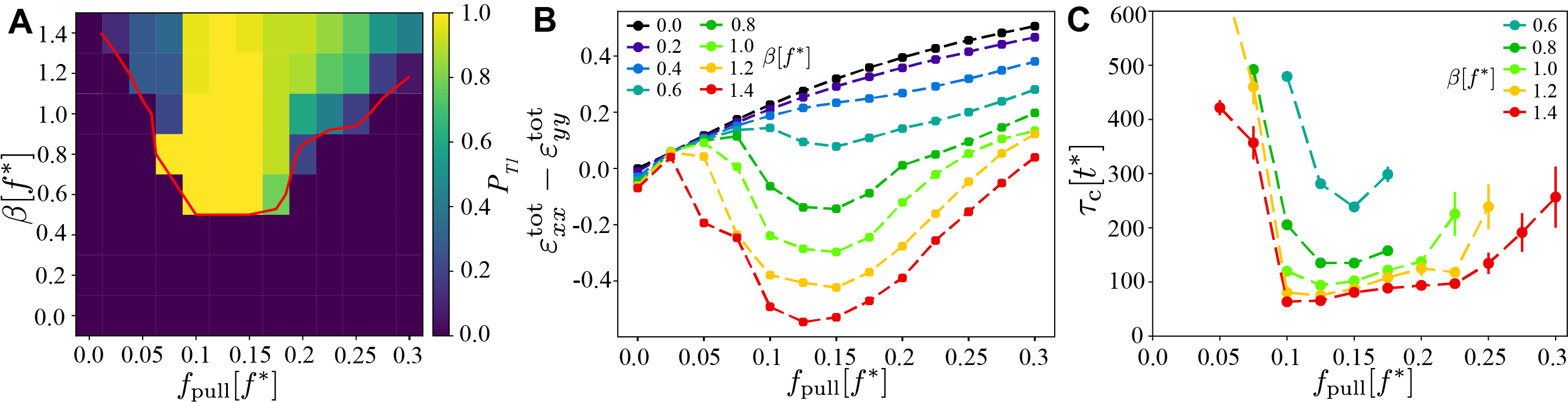}
    \caption{Existence and time scales of T1 transitions in the vertex model with active junctions as a function of $f_\text{pull}$ and $\beta$, averaged over $N=32$ simulations with different realisations of the myosin noise. \textbf{A}) Probability of a central T1 transition. The red line is the $50\%$ probability contour of any T1 occurring in the simulation. \textbf{B}) Magnitude of the convergence-extension deformation as a function of $f_\text{pull}$ and characterised by measuring $\varepsilon^\text{tot}_{xx}-\varepsilon^\text{tot}_{yy}$ induced by the T1 transition. \textbf{C}) Typical timescale for the T1 transition to occur, measured as length of the contraction phase. The other parameters are the same as in fig.~\ref{fig:T1_schematic}.  }
    \label{fig:T1_diag}
\end{figure*}

To understand dynamics of a single active T1, we first studied a regular lattice of hexagonal cells that are passive except for an inclusion of four central active cells surrounded by a buffer ring at half activity (fig.~\ref{fig:T1_schematic}A). The buffer cells were used to prevent distortions associated with large differences in activity between cells. Mechanical polarisation was created by applying forces of equal magnitude and opposite direction perpendicular to the left and right boundaries with both activity and viscoelastic relaxation switched off, until mechanical equilibrium was reached (see \emph{Materials and Methods}). As shown in fig.~\ref{fig:T1_schematic}C (top left), the initial state is mechanically polarised, with differential tension in horizontal (h) vs. shoulder (s) junctions (fig.~\ref{fig:T1_dynamics}A), with $T_\text{h}$ being significantly larger than $T_\text{s}$. Near the equilibrium point $T=T^*$, and for $M=6$, $m^\text{C}_\text{act}\approx3$, the dynamics of the model tissue closely resembles the dynamics of the single junction model. It is easy to show that $m_\text{eq} \approx \left(2 F(T)\right)^{-1}$, and mechanical polarisation leads to myosin polarisation, i.e. $m_\text{h} > m_\text{s}$ (fig.~\ref{fig:T1_schematic}C, bottom left). There is a range of applied pulling forces that, therefore, produce tensions $T_\text{s} < T^* < T_\text{h}$ in the system. In this regime, for a suitable $\beta> \beta_\text{c}/2$ (cf.~eqn.~(\ref{eq:beta_c})) horizontal junctions are contractile, while shoulder junctions are extensile. Here the factor of $1/2$ is due to both active elements of a junction acting in parallel.

In steady state, both activity $\beta$ and viscoelasticity were switched on (fig.~\ref{fig:T1_schematic}C and Movies S1-S6 in Supplementary Information), with time scales chosen such that $\tau_\text{v},\tau_\text{m}\gtrsim t^*$, which is the biologically relevant regime. The second column in fig.~\ref{fig:T1_schematic}C shows the system after $75t^*$. For the central horizontal active junction, contractility has been triggered, and the junction is steadily contracting at high myosin and against high tension, as shown in fig.~\ref{fig:T1_dynamics}B. 

The active T1 transition is reached at $121t^*$, when the central junction shrinks to a point and a four-vertex is created (third column in fig.~\ref{fig:T1_schematic}C). If the sum of the forces on the four-vertex is favourable for it to split \cite{spencer2017vertex} in the vertical direction, a T1 event occurs, accompanied by myosin redistribution (see Supplementary Information). The newly created vertical junction expands at intermediate values of myosin and tension, as shown in the fourth column in fig.~\ref{fig:T1_schematic}C and in fig.~\ref{fig:T1_dynamics}B. In contrast, aided by the redistribution of myosin, the shoulder junctions are now strongly polarised and begin contracting, with higher myosin levels on the side of the junction belonging to the expanding pair of cells (fig.~\ref{fig:T1_dynamics}C). This expansion phase is followed by several secondary T1 events, the first of which typically occurs at a shoulder junction (fifth column in fig.~\ref{fig:T1_schematic}C). During this propagation phase, the system remains strongly mechanically polarised in the direction of applied pulling forces. Together, the central T1 and the subsequent T1 events lead to substantial convergence-extension flow as can be seen qualitatively in the shape of the region formed by the 14 central and buffer cells (medium and dark grey in fig.~\ref{fig:T1_schematic}).

{\bf Time scales of active T1 events and local convergence-extension strain.} We proceed to quantify T1 transitions using the method introduced by Graner, et al.\ \cite{graner2008discrete} and summarised in Sec.~\ref{sec:Characterisation-of-the-T1-transition} in Supplementary Information. First, using the topological tensor, $\hat{\mathbf{T}}$ defined in eqn.~(\ref{eq:Ttensor}), we measured the time and orientation of the T1 transition along the direction the central junction of the active region (hereafter the ``central T1''). Figure~\ref{fig:T1_diag}A shows the probability of a central T1 as a function of $f_\text{pull}$ and $\beta$, with other parameters held constant at the same values as in fig.~\ref{fig:T1_schematic}. The probability was computed from $n=32$ simulations with different realisations of the myosin noise as the fraction of simulations where the first observed T1 was along the central active junction rather than elsewhere in the system. The probability of any T1 in the system was also measured, with the red contour in fig.~\ref{fig:T1_diag}A corresponding to $50\%$ of realisations having a T1.

These results show that there is an absolute lower threshold, $\beta > \beta_\text{c} = 0.6f^*$ for any form of T1 to occur. This is qualitatively consistent with both sides of the junction acting as two parallel instances of the single junction model. Second, there is an optimal range of applied pulling forces for central T1 transitions, $0.1<f_\text{pull}/f^*<0.2$. Within this range, $T_\text{s} < T^* < T_\text{h}$ for the central and shoulder junctions, and during the initial contracting phase, they are contractile and extensile, respectively. Outside this optimal regime, the probability for central T1 events decreases rapidly, though T1 transitions still occur elsewhere for large values of $\beta$. 

\begin{figure*}[tb]
    \centering
    \includegraphics[width=0.99\textwidth]{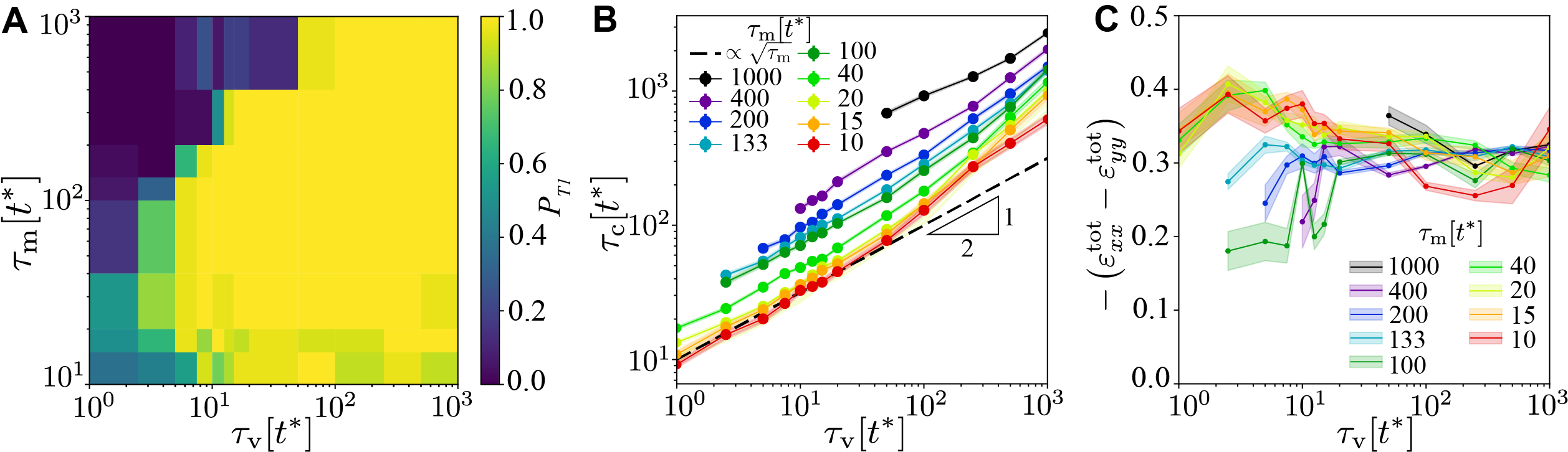}
    \caption{Robustness of the T1 mechanism as a function of $\tau_\text{v}$ and $\tau_\text{m}$ for $\beta=1.0f^*$ and $f_\text{pull}=0.15f^*$. \textbf{A}) Probability of a central T1 event, averaged over $n=32$ simulations with different realisations of the myosin noise. The probability of any T1 event is $1$ throughout. \textbf{B}) Contraction time to collapse for the central T1 as a function of $\tau_\text{v}$, for different values of $\tau_\text{m}$, for points where a central T1 event occurred in at least $25\%$ of simulations. \textbf{C}) Peak of the total convergence extension strain $\varepsilon^\text{tot}_\text{xx}-\varepsilon^\text{tot}_\text{yy}$, showing very weak dependence on viscoelastic and myosin time scales. Shading in panels B and C indicates the standard error of the mean.}
    \label{fig:T1_times}
\end{figure*}

The orientation of the central T1 transition in the optimal region is shown in fig.~\ref{fig:T1_dynamics}D. It was computed from the angle of the principal direction of $\hat{\mathbf{T}}$ corresponding to the largest eigenvalue before and after the T1. One can immediately observe that the transition is highly symmetric, with the orientation of the shrinking (disappearing) junction being very close to horizontal, and the expanding (appearing) junction very close to vertical. Outside the optimal regime, this symmetry disappears, with T1 events of non-central junctions occurring in different directions. 

Figure~\ref{fig:T1_diag}C shows the time to the first $T1$ transition, $\tau_\text{c}$, measured from the point when the activity and viscoelastic relaxation were switched on. The analysis was limited to central T1 transitions at parameter values where at least $25\%$ of simulations yield a central T1, with $\tau_\text{v}=20t^*$ and $\tau_\text{m}=100t^*$. One immediately observes that $\tau_\text{c}\gtrsim \tau_\text{v},\tau_\text{m}$, consistent with a T1 dynamics being dominated by myosin activation and viscoelastic relaxation. We find that $\tau_\text{c}$ has a minimum in the same optimal region identified in fig.~\ref{fig:T1_diag}A, with $\tau_\text{c}$ rising both for larger and smaller values of $f_\text{pull}$. Furthermore, increasing $\beta$ beyond $\beta_\text{c}=0.6f^*$ gradually reduces $\tau_\text{c}$, consistent with active contractions becoming stronger. For $\beta>1.0f^*$, cells shapes become increasingly distorted, suggesting that the model is no longer applicable. 

We now quantify convergence-extension generated by the model. As outlined in Sec.~\ref{sec:Characterisation-of-the-T1-transition} in Supplementary Information, the shear component of the total integrated strain $\varepsilon^\text{tot}_\text{xx}-\varepsilon^\text{tot}_\text{yy}$ given in eqn.~(\ref{eq:tot_strain}), was computed for the 14 cells comprising the central and buffer regions (fig.~\ref{fig:T1_schematic}A). Figure~\ref{fig:T1_diag}B shows $\varepsilon^\text{tot}_\text{xx}-\varepsilon^\text{tot}_\text{yy}$ as a function of $f_\text{pull}$, evaluated at $t=400t^*$ and averaged over $10t^*$. This point approximately corresponds to the empirically determined peak of convergence-extension in the optimal region of applied pulling forces (see sample time traces in fig. \ref{fig:Graner_traces} in Supplementary Information). From fig.~\ref{fig:T1_diag}B it is evident that the model generates pronounced convergence-extension. Without activity, i.e. for $\beta=0$, the system extends in the direction of the applied pulling force and contracts perpendicular to it with $\varepsilon^\text{tot}_\text{xx}-\varepsilon^\text{tot}_\text{yy}>0$ and, as expected, it increases with $f_\text{pull}$. For $\beta>0$, $\varepsilon^\text{tot}_\text{xx}-\varepsilon^\text{tot}_\text{yy}>0$ decreases, indicating the activity acts against the applied pulling force. As $\beta$ increases beyond a critical value, $\beta_\text{c}\approx0.6f^*$, active forces are strong enough to counteract the pulling forces and the system shrinks against the external load and extends in the perpendicular direction, i.e. $\varepsilon^\text{tot}_\text{xx}-\varepsilon^\text{tot}_\text{yy}<0$. During this process there are no significant changes of the area, i.e. $\varepsilon^\text{tot}_\text{xx}+\varepsilon^\text{tot}_\text{yy}\approx0$.
This mechanism is, however, effective only for a range of values of $f_\text{pull}$. If $f_\text{pull}$ is insufficiently strong, the myosin-tension feedback loop does not fully activate (Movie S4 in Supplementary Information). Conversely, if $f_\text{pull}$ is too strong, the feedback loop is active, but all junctions are activated, stiffening the tissue (Movie S5 in Supplementary Information). 

\begin{figure}[tb]
    \centering
    \includegraphics[width=0.99\columnwidth]{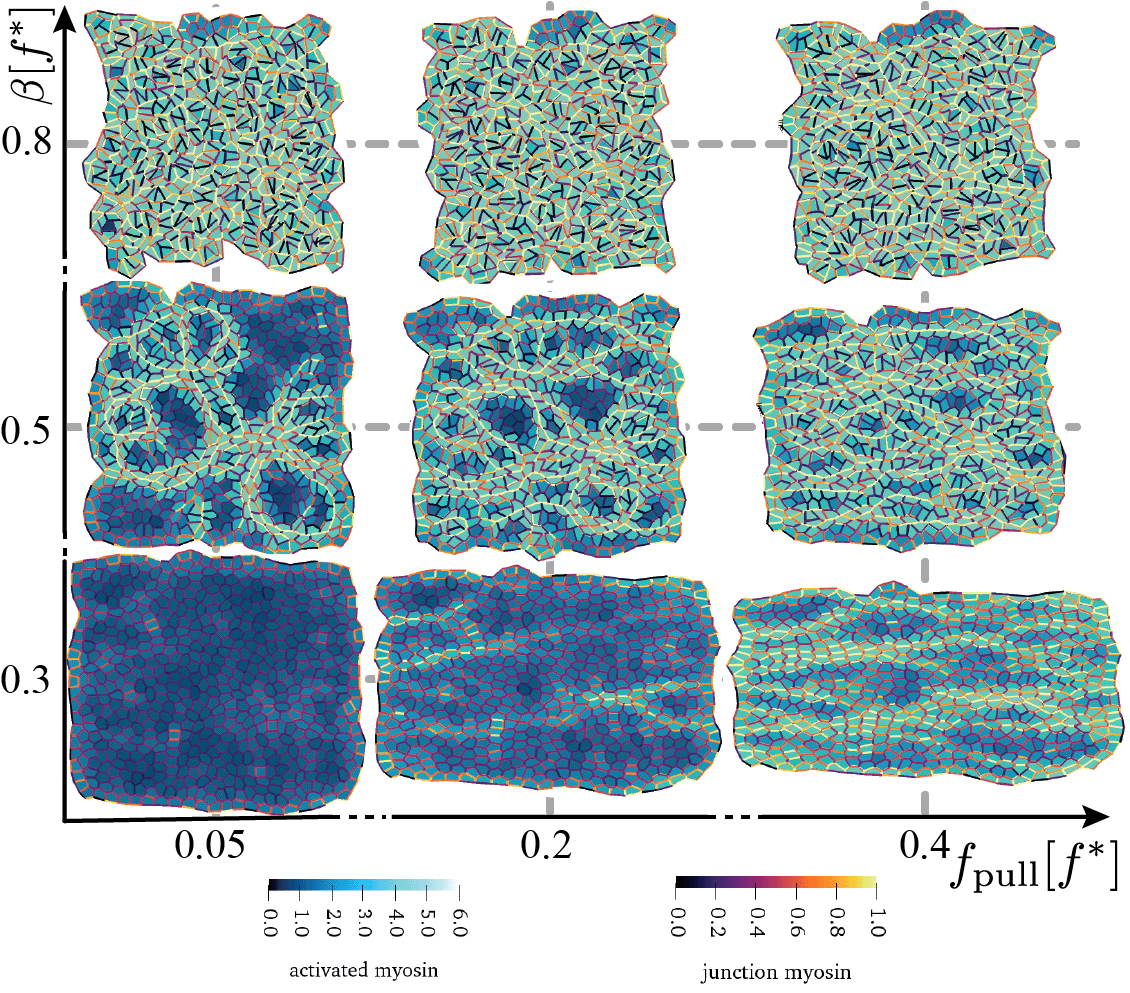}
    \caption{Disordered active tissue at time $t\approx 700t^*$ as function of the magnitude of the pulling force $f_\text{pull}$ and activity $\beta$. The region of convergence-extension is at the centre of the diagram, around $\beta \approx 0.4 - 0.6f^*$ and $f_\text{pull}\approx 0.1-0.3f^*$. The remaining parameters are the same as in fig.~\ref{fig:T1_schematic}.}
    \label{fig:random_phase_diagram}
\end{figure}

We conclude the analysis of a hexagonal tissue patch by investigating the influence of $\tau_\text{v}$ and $\tau_\text{m}$ (see Movies S7-S13 in Supplementary Information). Figure~\ref{fig:T1_times}A shows the probability of a central T1 for $\beta=1.0f^*$, $f_\text{pull}=0.15f^*$, i.e. deep in the optimal region, as a function of $\tau_\text{m}$ and $\tau_\text{v}$. Here, we only consider the biologically plausible regime with $\tau_\text{m}, \tau_\text{v} \gtrsim t^*$, with proportionally scaled myosin noise $f=1$, and we excluded very small values of $\tau_\text{m}$ where noise dominates. We find that the mechanism for T1 events is very robust over 2--3 orders of magnitude in both $\tau_\text{m}$ and $\tau_\text{v}$, with a guaranteed T1 transition in most of the parameter space. The only exception is the regime $\tau_\text{m} \gtrsim 10\tau_\text{v}$, i.e. very slow myosin dynamics compared to viscous relaxation, where the system fails to polarise. All simulations generated at least one T1, and there is no equivalent of the red contour in fig.~\ref{fig:T1_diag}A. In fig.~\ref{fig:T1_times}B, we show the timescale of the T1 transition. We find that the T1 time scales as $\tau_\text{c} \sim \tau_\text{v}^{1/2}$ and $\tau_\text{c} \sim \tau_\text{m}^{1/3}$. This influence of both time scales is consistent with the complex interplay between myosin activation on central and shoulder junctions.
Finally, in fig.~\ref{fig:T1_times}C, we show the convergence-extension strain $\varepsilon^\text{tot}_\text{xx} - \varepsilon^\text{tot}_\text{yy}$ as a function of $\tau_\text{v}$ for a range of values of $\tau_\text{m}$. The effectiveness of the T1 mechanism is largely independent of $\tau_\text{v}$ and $\tau_\text{m}$, and we have $\varepsilon^\text{tot}_\text{xx} - \varepsilon^\text{tot}_\text{yy} \approx -0.3$, the same value as in fig.~\ref{fig:T1_diag}B. This additional robustness of the model to varying time scales is likely due to being in a quasistatic regime where the elastic deformation timescale is much shorter than any other timescale.

\begin{figure*}[bt]
    \centering
    \includegraphics[width=0.99\textwidth]{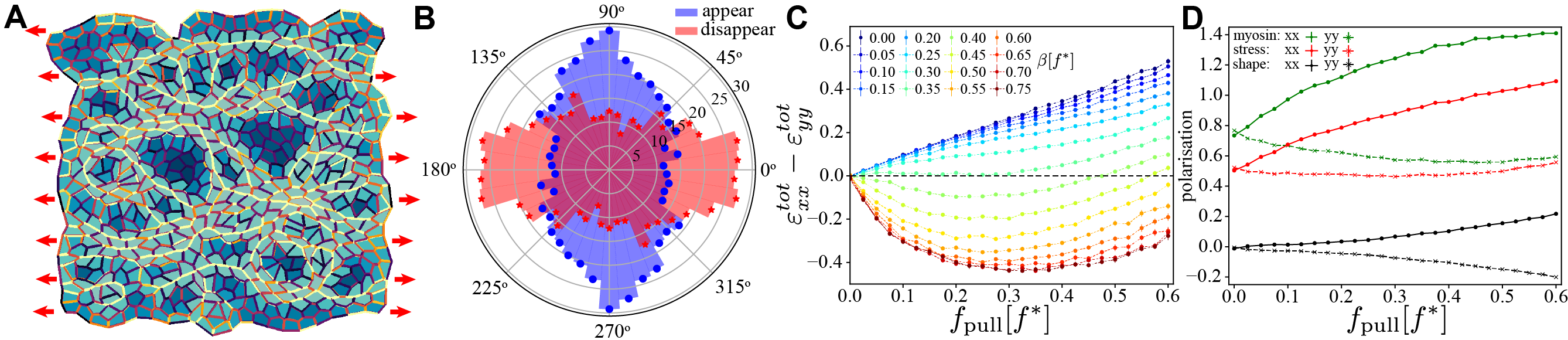}
    \caption{{\bf A}) Snapshot of a random tissue patch for $\beta=0.5f^*$ and $f_\text{pull}=0.2f^*$ at $t\approx700t^*$, i.e. during the convergence-extension flow. Red arrow indicate that a constant pulling force is applied throughout the entire simulation. {\bf B}) Angular histogram of T1 events for same values of $\beta$ and $f_\text{pull}$. Cells and junctions are coloured as in fig.~\ref{fig:random_phase_diagram}. {\bf C})  Magnitude of the convergence-extension deformation as a function of $f_\text{pull}$ characterised by measuring $\epsilon^\text{tot}_{xx}-\epsilon^\text{tot}_{yy}$ induced by the T1 transition. {\bf D}) Polarisation along ($xx$, solid line) and perpendicular to ($yy$, dashed line) the direction of the external pulling force for myosin (green), mechanical stress (red) and shape tensor (black) as functions of $f_\text{pull}$ for $\beta=0.5f^*$ at $t\approx700t^*$. In C and D, each point was averaged over $n=33$ independent samples and the error bar is smaller than the symbol size. }
    \label{fig:random_patches}
\end{figure*}

{\bf Convergence-extension in a fully active random patch.} The hexagonal tissue patch is convenient to analyse isolated active T1 events. Cells in real epithelia, however, do not have regular shapes packed in crystalline order. We, therefore, studied a patch of $520$ active and $80$ passive cells generated from a centroidal Voronoi tessellation starting from $N=600$ points placed at random in the simulation box. First, we investigated the occurrence of active T1 transitions and the emergence of convergence-extension as a function of $f_\text{pull}$ and $\beta$, at fixed $\tau_\text{v}=20t^*$ and $\tau_\text{m}=100t^*$, i.e. in the same region of parameter space as in fig.~\ref{fig:T1_diag} (see also fig.~\ref{fig:random_phase_diagram}). To be consistent with the hexagonal patch, the unit of length here is set by the length of a regular hexagon of area $L^2/N$, where $L$ is the initial patch size (see \emph{Materials and Methods}).

We observed T1 transitions for all simulated systems, though their rate rapidly increased when either $\beta$ or $f_\text{pull}$ were increased. Unlike in the case of the active inclusion in a hexagonal passive patch, there is no clear threshold for active T1 transitions. Instead, parts of the tissue are activated, and we observed the emergence of pronounced myosin cables and accompanying tension chains (e.g. middle row in fig.~\ref{fig:random_phase_diagram}). The system starts to experience significant flow and convergence-extension from $\beta = 0.4f^*$, significantly below the values observed in the hexagonal patch with a single active inclusion where, depending on the magnitude of the applied force, T1 events start to appear for $\beta$ above $0.6-1.0f^*$. This suggests cooperative rearrangements in the tissue, and we indeed see evidence of serial active T1 transitions along tension chains (see Movie S14 in Supporting Information). This suggests that the individual junction feedback mechanism together with mechanical (as opposed to chemical) propagation of myosin activation is a key ingredient in the formation of the myosin cables that have been observed to accompany convergence-extension flow in chick embryo gastrulation \cite{rozbicki2015myosin} and in Drosophila germ band extension \cite{jacinto2002dynamic}.

The system starts to rearrange from $\beta \approx 0.4f^*$, and for $\beta \gtrsim 0.7f^*$, one observes a highly active state with many uncorrelated rearrangements and implausible cell shapes. The region with realistic cell shapes is significantly below the hexagonal patch with a single active inclusion, where, depending on the magnitude of the applied force, T1 events start to appear for $\beta$ above $0.6-1.0f^*$.

In fig.~\ref{fig:random_patches}A we show a snapshot of a random patch long after ($\approx700t^*$) activity was switched on for $\beta=0.5f^*$ and $f_\text{pull}=0.2f^*$, in what emerges to be the optimal region for convergence-extension. Oriented myosin cables accompanied by tension chains form in the initial stage of the simulation at $\approx 100t^*$ after activity was turned on. At longer times, the pronounced orientation of myosin and tension decreases (but does not disappear entirely), and convergence-extension stops (see Sec.~\ref{Sec:SI-CE-random} in Supplementary Information for details). The convergence-extension process is accompanied by active T1 events that predominately occur perpendicular to the direction of the external pulling, as shown in the orientational histogram in fig.~\ref{fig:random_patches}B. 

\begin{figure*}
    \centering
    \includegraphics[width=0.85\textwidth]{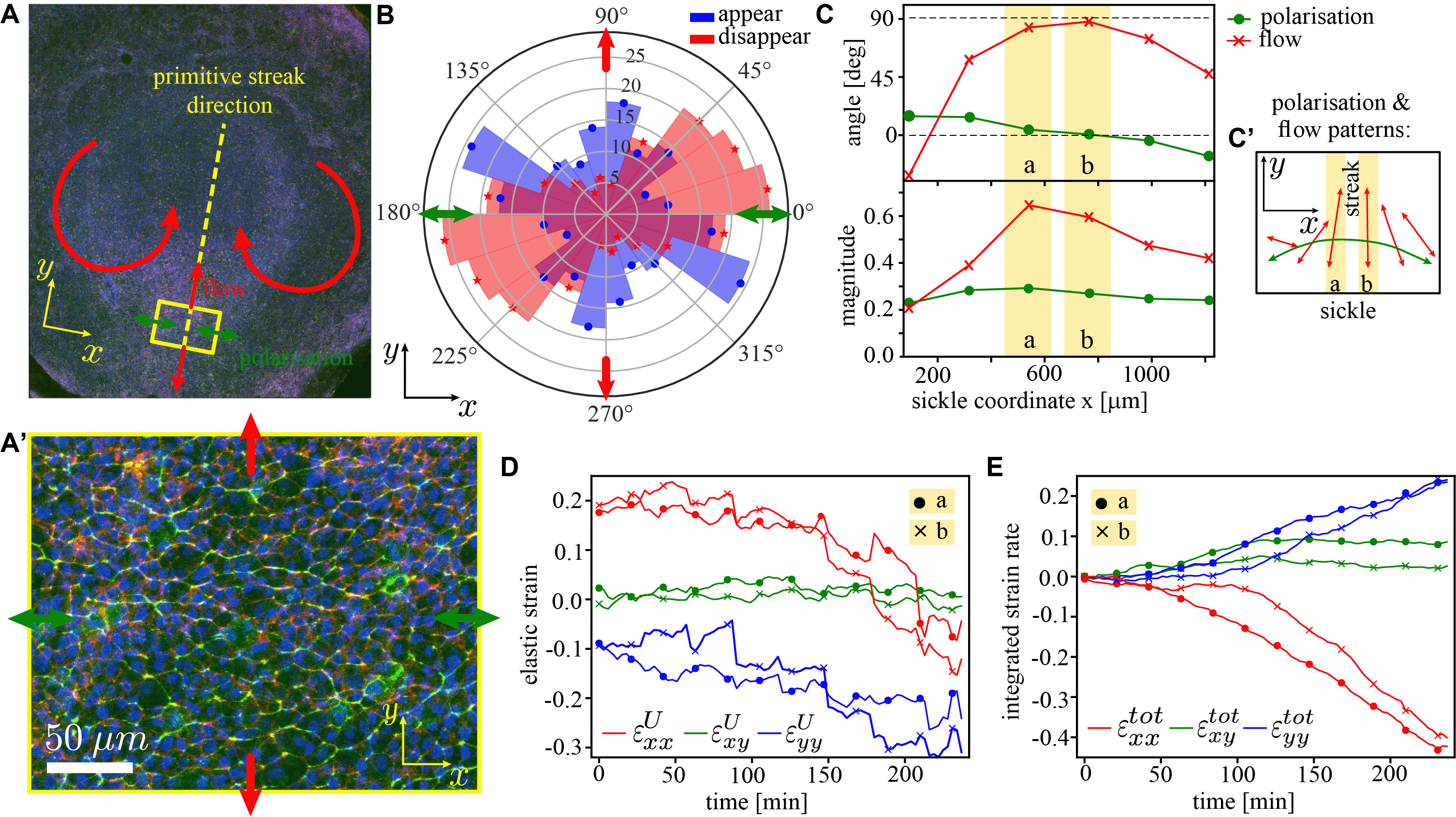}
    \caption{Analysis of the tissue flows in the early-stage chick embryo. {\bf A}) Image of a typical early-stage chick embryo prior to the gastrulation (i.e. primitive streak formation). The primitive streak will form along the yellow dashed line. The direction of myosin polarisation is shown by green double-headed arrows and the direction of the tissue flow is indicated by read arrows. $x-$axis is chosen to coincide with the long direction of the sickle-shaped active region in the embryo's posterior \cite{rozbicki2015myosin}. {\bf A'}) Zoom-in of the rectangular region on the posterior side of the embryo; myosin-II (green), actin (red), and nuclei (blue). {\bf B}) Measured distribution of the orientation of T1 events in a circular patch of diameter $\approx190$ $\mu\text{m}$ tracked over the period of $\approx6$ h (cf.~model distribution in fig.~\ref{fig:random_patches}B). Blue (red) denoted junctions that appear (disappear); arrows have the same meaning as in {\bf A}. {\bf C}) Angle (red) and magnitude (green) of tissue shape polarisation (dots) and tissue flow (crosses) for $n=6$ rectangular patches along the sickle with corresponding polarisation and flow patterns shown in {\bf C'}. Components of the elastic strain tensor $\hat{\mathbf{U}}$ (panel {\bf D}) and the total integrated strain tensor $\hat{\mathbf{V}}$ defined in eqn. (\ref{eq:tot_strain}) (panel {\bf E}) as a function of time during first 4h of the streak formation for two central regions of the sickle (yellow stripes in {\bf C}). Details of the analysis are given in Sec.~\ref{sec:SI-experiment-data-analysis} in Supplementary Information.}
    \label{fig:experiment}
\end{figure*}

Figure \ref{fig:random_patches}C quantifies the amount of convergence-extension by measuring the difference of total integrated strain in directions along and perpendicular to the direction of the applied pulling force, i.e. $\epsilon^\text{tot}_{xx}-\epsilon^\text{tot}_{yy}$. It is evident that like the hexagonal case, the random patch undergoes substantial convergence-extension over a range of activities. The process in accompanied by spatial polarisation of myosin, mechanical stress and cell shapes (fig.~\ref{fig:random_patches}D), which we measured through the eigenvalues of the myosin and tension tensors as, respectively defined in eqns.~(\ref{eq:SI-M_myo}) and (\ref{eq:SI-T_tension}), and the eigenvalues of the shape tensor, eqn.~(\ref{eq:Utensor})) in Supplementary Information.

{\bf Comparison with experiments on gastrulation in early-stage chick embryo.} We make a qualitative comparison of the model with tissue deformations observed in early-stage chick embryos. Prior to the formation of the primitive streak (i.e. the initial stage of gastrulation in avian embryos), the \emph{Gdf3} growth factor is expressed in the sickle shaped region located at the embryo's posterior, perpendicular to the direction of the streak formation \cite{najera2020cellular}. This is accompanied by myosin polarisation along the sickle-region \cite{rozbicki2015myosin}. The sickle region develops into the primitive streak via large-scale tissue deformations characterised by pronounced convergence-extension flows \cite{rozbicki2015myosin}. Figure \ref{fig:experiment}A shows an image of a typical embryo prior to primitive streak formation with a patch magnified in panel A'. Both myosin (green) and cell shapes are polarised along the left-right axis (i.e. along the sickle) as indicated by green arrows. The tissue flows in the perpendicular direction, indicated by red arrows. One can observe pronounced myosin cables and that are believed to generate tension chains, as observed in simulations. 

We quantified convergence-extension as follows. Figure \ref{fig:experiment}B shows an orientational histogram of T1 events in a rectangular region of size $\approx 200\times200\ \mu \text{m}^2$ along the sickle tracked over a period of approximately 6 h. The events were identified by calculating the $\hat{\mathbf{T}}$ tensor using segmented images (see Sec.~\ref{sec:SI-experiment-data-analysis} and Movie S15 in Supplementary Information), and validated and corrected manually. In fig.~\ref{fig:experiment}C we analysed $n=6$ tissue patches of diameter $\approx 190$ $\mu$m chosen sequentially along the sickle-shaped region. We computed tensors that measure shape polarisation $\hat{\mathbf{U}}$ and strain rate $\hat{\mathbf{V}}$ \cite{graner2008discrete} by tracking the patches along the tissue flow over approximately 4 h (see Sec.~\ref{sec:SI-experiment-data-analysis} in Supplementary Information). The polarisation remains constant at around $20\%$ and the polarisation angle remains close to $90^{\circ}$, i.e. along the sickle and orthogonal to the streak (green arrow in fig.~\ref{fig:experiment}C'). We also computed the mean flow magnitude and direction from the eigenvalues and eigenvectors of the integrated $\hat{\mathbf{V}}$ tensor. There is a pronounced spatial pattern to its direction, pointing towards the streak and parallel to the polarisation on outer parts of the sickle, and orthogonal to the sickle and its polarisation in the middle of the sickle (red arrows in fig.~\ref{fig:experiment}C'). At the same time, the magnitude of flow peaks in the middle part of the sickle. This is consistent with the incipient flow to create the streak. In fig.~\ref{fig:experiment}D\&E, we show the time dependence of polarisation and total strain, for two central patches. We see that polarisation is along $x$ (i.e. perpendicular to the streak) and flow is along $y$ (i.e. along the streak), corresponding to convergence-extension flow. In fig.~\ref{fig:UV_anterior} and Movie S16 in the Supplementary Information, for comparison, we show the behaviour in a non-polarised region anterior to the streak. 

\section*{Discussion}

There are two key features of our model with active junctions. First, myosin polarity is induced by anisotropic mechanical tension. Second, active contractions are triggered by tension-sensitive accumulation. The feedback loop between tension and myosin motor activity leads to contraction against and extension perpendicular to tension. This cellular mechanism has been suggested to be driving the tissue flows during primitive streak formation in avian embryos \cite{rozbicki2015myosin}, where there is no clear evidence of chemical prepatterning. Although it is yet to be experimentally confirmed, it is plausible that the symmetry breaking event that induces the initial myosin polarity occurs as a result of anisotropic tension combined with cell differentiation early in development. For example, in the chick embryo, \emph{Gdf3} is expressed in the sickle-shaped region in the posterior epiblast, and it is believed to play a role in triggering the contractions that initiate the large scale tissue flows that subsequently lead to formation of the primitive streak \cite{najera2020cellular}. The key conclusion of this study is that once the process has been initiated, chemical polarisation emerges spontaneously and there is no need to impose it. 

Although the model investigated here generates active T1 events, the process loses coherence after several T1 events resulting in biologically implausible tissue shapes. This is very prominent in the toy case of regular hexagonal patch. While the problem is to some extent alleviated in patches of randomly shaped active cells, generating robust convergence-extension flows that would span scales of the entire embryo will be hard to achieve with the current model. One source of instability is likely that the post-T1 expansion of the junction, currently effectively modelled as a passive process, is not properly captured by the model, and requires additional sources of activity to be considered. At the scale of the entire embryo, other cellular processes such as cell division, differentiation, and ingression all play non-trivial roles. These events have not been considered here.

Furthermore, recent results confirm that the mechanics of vertex models is complex \cite{bi2016motility,yan2019multicellular,tong2021linear}, which makes the tissue-scale flows that emerges from active elements coupled to the vertex model hard to predict using continuum approaches. Active nematic flows have been observed and modelled in in-vitro epithelial tissues~\cite{saw2017topological}, and due to the locally contractile and extensile dynamics it is plausible that large-scale flows predicted by this model belong to the same class of models. In real tissues, however, the mechanisms that regulate how cells coordinate their internal mechanical stress directions in order to produce a stable flow pattern are unknown. Active vertex models therefore provide valuable new insights into the intricate interplay between mechanical and biochemical processes that control the collective cell behaviours in epithelial tissues.

We also briefly discuss how this work relates to other recent models for active junction contractions and convergence extension. The single junction model shares a lot of common features with the model of Dierks, et al.~\cite{dierkes2014spontaneous}. The key difference, however, is a different myosin-tension feedback mechanism and that the dashpot in \cite{dierkes2014spontaneous} was replaced with a Maxwell element. This was inspired by laser tweezer measurements of the response of cell-cell junctions in the Drosophila embryo to applied pulling force~\cite{clement2017viscoelastic} which showed that the cellular junctions behave as a Maxwell viscoelastic material. The presence of an elastic spring attached to the dashpot introduces a viscoelastic timescale, which leads to the suppression of the oscillatory behaviour seen in~\cite{dierkes2014spontaneous}.

The model of Staddon, et al.~\cite{staddon2019mechanosensitive} considered a Maxwell element subject to active contraction, with the spring constant of the cellular junctions that constantly remodels itself to match strain in the junctions. The tension, however, remodels only if the strain exceeds a threshold value. These two features combine to provide a simple mechanism by which the junction can undergo ratchet-like behaviour and contract to a T1 event, and where T1 events can be triggered by applying external forces. Ref.~\cite{saadaoui2020tensile}, however, does not explicitly include kinetic equations for molecular motors.

The two-dimensional models of Wang, et al.~\cite{wang2012cell} and Lan, et al.~\cite{lan2015biomechanical} provide detailed descriptions for coupling between chemical signalling and corresponding mechanical responses. While they were able to produce T1 events, chemical polarity in the cell was externally imposed by tuning concentrations of relevant molecular species based on the origination of the junctions. This was also the case for the two-dimensional version of the model of Staddon, et al.~\cite{staddon2019mechanosensitive}, which was able to produce convergence-extension flows, albeit with time-dependent activity imposed in a given direction. Meanwhile, Noll, et al.~\cite{noll2017active} have also introduced a vertex model with junctions that incorporate generic active feedback in a model for tissue contraction in Drosophila gastrulation. They did not, however, consider active T1 transitions.

In summary, in this study we have introduced a mechanochemical model that describes the dynamics of active T1 transforms, i.e. cell intercalation events that occur perpendicular to the externally applied mechanical stress. Such processes are believed to play a key role in the primitive streak formation in avian embryos. Crucially, this study suggests that mechanical propagation of activation of myosin is a key ingredient in formation of the myosin tension chains that have been observed to accompany convergence-extension flow in chick embryo gastrulation \cite{rozbicki2015myosin} and in Drosophila germ band extension \cite{jacinto2002dynamic}. Finally, results of this study show a good qualitative agreement with measurements on early stage chick embryos. 

\section*{Materials and Methods}

See Supplementary Information, Sec.~\ref{subsec:SI-single-junction} for details of the single junction model, and Sec.~\ref{sec:SI-Vertex-Model-with-Active-Junctions} in Supplementary Information for the details of full two-dimensional model.

The equations of motion for the two-dimensional model were integrated numerically for a rectangular patch made of $N=158$ hexagonal cells and patch made of $N=600$ randomly shaped cells using open boundary conditions. Mechanical polarisation was created by applying a force $\mathbf{f}_\text{pull}=\pm f_\text{pull}\mathbf{e}_\text{x}$ to the left and right boundary vertices (fig.\ \ref{fig:T1_schematic}A), where the positive (negative) sign corresponds to the right (left) boundary. The initial pull was applied for $10^3t^*$ with both activity and viscoelastic relaxation switched off, sufficient to reach mechanical equilibrium. Once the system reached an equilibrium stretched state, activity was switched on in $14$ central cells in the case of the hexagonal patch and for $520$ cells in the case of the random tissue patch. For the hexagonal case, activity was set to $\beta$ in $4$ cells and to $\beta/2$ in $10$ ``buffer'' cells surrounding them, in order to suppress numerical instabilities at contacts between active and passive cells. For the random patch, activity was set to $\beta$ in all cells except for a single-cell thick layer of boundary cells that were kept passive to prevent artefacts due to tension chains reaching the sample boundary. An external pulling force of constant magnitude was applied throughout the entire simulation. The active system was simulated for $\max\left(1600t^*,10\tau_\text{m},10\tau_\text{v}\right)$ using time step $10^{-2}t^*$. In the hexagonal case, the dynamics of the central horizontal junction and the shoulder junctions shared by the four central active cells was monitored. The orientation of the hexagonal lattice was chosen such that the central active junction was parallel to the direction of the applied external force.  In the case of the random patch, the dynamics of all junctions shared by active cells was monitored. 

The T1 events were implemented following the procedure proposed by Spencer, et al.\ \cite{spencer2017vertex}, where a junction shorter than $0.02a$ collapses into a four-fold vertex. The four-fold vertex either remains stable or it is resolved into two three-fold vertices based on the sum of the forces acting along the four junctions connected to it. Importantly, the direction of the new junction is not imposed and this procedure does not generally lead to a new junction orthogonal to the collapsed one. For simplicity, vertices with connectivity greater than four were not considered.

Finally, if a T1 transition occurs it is necessary to assign myosin to the newly created junction and to redistribute the myosin associated to the collapsed junction to the surrounding junctions. As illustrated in and using the notation introduced in fig.~\ref{fig:AJM_schematic}B in Supplementary Information, the myosin $m_\text{e}$ of the collapsing junction is stored right before the junction collapses. After the T1 transition, the myosin on the new junction was set to $m_0$. Both inner shoulder junctions increased their myosin by $m_\text{e}/2$, while the myosin on the outer junctions was reduced by $\min\left(m_2,m_0/2\right)$, where $m_2$ is its myosin right before the T1 transition and the minimum function ensures that myosin remains $\ge0$. The myosin redistribution procedure reduces artificial jumps of the junctional myosin as the system progresses through T1 transition and also provides a natural way to handle the finite myosin pool. After the transition, $l_0$ of the new junction is set to $0.022a$. 

\section*{Acknowledgments}

R.S. and C.J.W.~acknowledge support by the UK BBSRC (Award BB/N009789/1). S.H.~and I.D.C.~acknowledges support by the UK BBSRC (grant number BB/N009150/1-2). I.D.C.  acknowledges funding under Dioscuri, a programme initiated by the Max Planck Society, jointly managed with the National Science Centre in Poland, and mutually funded by Polish Ministry of Science and Higher Education and German Federal Ministry of Education and Research (UMO-2019/02/H/NZ6/00003). We thank Antti Karjalainen for providing the MATLAB code used to analyse experimental data. R.S. thanks Andrej Ko{\v s}mrlj, Daniel Matoz-Fernandez, and Sijie Tong for many helpful discussions about the vertex model.

\section*{Data availability} The data supporting the results and findings of this study is available from the corresponding authors upon reasonable request.

\section*{Code availability} Simulation and analysis codes used in this study are available from R.S. upon reasonable request.

\section*{Author Contributions}

C.J.W. formulated the problem. S.H., R.S, and C.J.W. developed the model. R.S. implemented the vertex model with active junctions. S.H., I.D.C., and R.S. carried out simulations and analysed the data. M.C. and C.J.W. performed experiments and analysed experimental data. R.S., I.D.C, C.J.W, and S.H. wrote the paper.

\section*{Competing interests}
The authors declare no competing interests.

\renewcommand{\thefigure}{S\arabic{figure}}
\renewcommand{\theequation}{S\arabic{equation}}
\renewcommand{\thetable}{S\arabic{table}}
\renewcommand{\thesection}{S\arabic{section}}
\setcounter{figure}{0} 
\setcounter{equation}{0} 
\setcounter{table}{0}
\setcounter{section}{0}
\setcounter{subsection}{0}

\clearpage
\newpage
\onecolumngrid

\begin{center} {\bf \large{Generating active T1 transitions through mechanochemical feedback}} \\
     \large{ Supplementary Information}
\end{center}

\section{Model setup and analysis}

\subsection{Single active junction} \label{subsec:SI-single-junction}

To understand the mechanism that couples the kinetics of myosin motors to the local mechanical tension and leads to the activation of contractility in cell-cell junctions, we first analyse a model for a single junction. The surrounding tissue is abstracted by assuming that it provides an elastic, tension-generating background against which the junction actively contracts. There are two key ingredients that make an active junction. First, the junction is viscoelastic. Pull-release optical tweezer experiments on cell-cell junctions in Drosophila and chick embryos have shown that cell-cell junctions have a viscoelastic response~\cite{clement2017viscoelastic,ferro2020measurement}. This means that the junction is able to remove imposed tension by remodelling itself. Second, the junction can generate tension. Tension is generated by myosin motors that form mini filaments slide actin filaments past each other. The single-junction model provides insight into the conditions under which the junction length can contract to zero and trigger a T1 transition. 

\begin{figure}
\includegraphics[width=\textwidth]{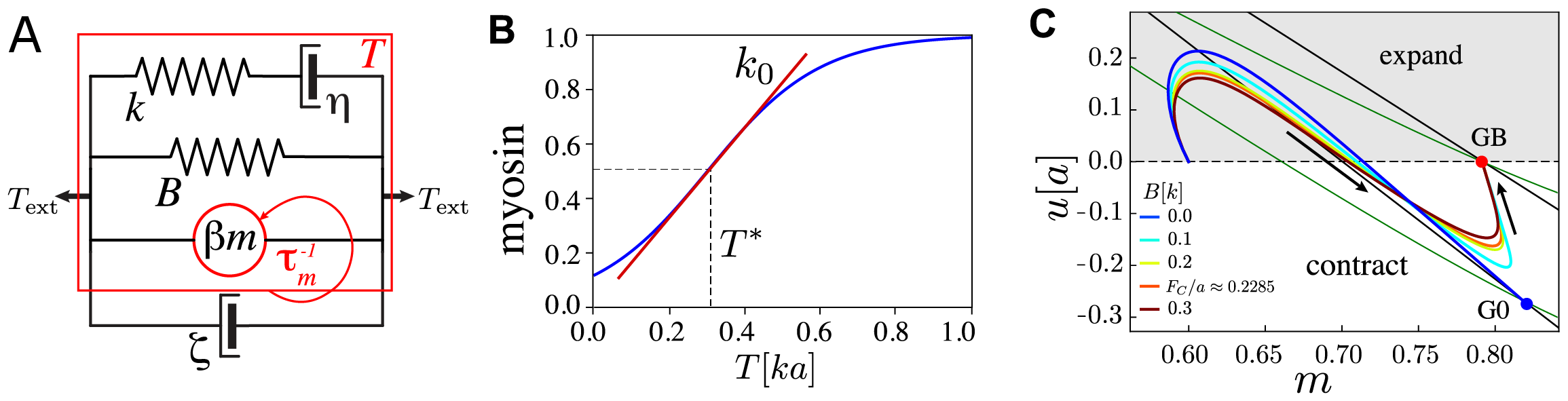}
\caption{\textbf{A}) The junction is modelled as a Maxwell element with stiffness $k$ and viscosity $\eta$ (and the relaxation timescale $\tau_\text{v}=\eta/k$) connected in parallel with the active forcing, and an elastic barrier with stiffness $B$ and rest length $a$. It is coupled in parallel to a dashpot with viscosity $\zeta$ modelling dissipation with the surrounding medium. The red arrow indicates the feedback loop between tension and myosin kinetics. \textbf{B}) Myosin-tension response curve, for $F(T) = \alpha + e^{-k_0\left(T-T^*\right)}$ with $\alpha=1$, $T^*=0.3ka$, $k_0=2/T^*$, $m_0 = 0.5$. \textbf{C}) Trajectories in the $u-m$ plane for several values of $B$, where $u=l-l_0$. G0 (GB) is the fixed point with no ($B>0$) elastic barrier. Fixed points are located at the intersection of their respective $u$ (black) and $m$ (green) nullclines. Arrows indicate the direction of motion.  $\alpha=1$, $T^*=0.3ka$, $k_0=2/T^*$, and $m_0=0.5$. Length is measured in units of $a$, time in units $t^*=\zeta/k$, and force in units of $ka$. }
\label{fig:SI_model}
\end{figure}

Specifically, the single junction (fig.~\ref{fig:junction_barrier}A in the main text) is modelled as an active mechanochemical system comprising three components connected in parallel (fig.~\ref{fig:SI_model}A): 1) a viscoelastic Standard Linear Solid (SLS) element, 2) a viscous dashpot, and 3) a tension-sensitive force-generating motor. The junction is subject to external tension $T_\text{ext}$ produced and transmitted by the surrounding cells. The SLS element consists of an elastic spring of stiffness $B$ and rest length $a$ connected in parallel with a Maxwell element containing a spring of stiffness $k$ and rest length $l_0$ attached in series to a dashpot of viscosity $\eta$~\cite{larson1999structure}. The spring $B$ captures the passive elastic response of the junction and models the effects of the surrounding tissue. In the two-dimensional model discussed below, this term arises naturally and accounts for changes in the cell area and perimeter. It is referred to as the \emph{elastic barrier}. The Maxwell element models the viscoelastic nature of the junction and the ratio of viscosity and stiffness sets its relaxation timescale $\tau_\text{v} = \eta/k$. This is the timescale over which the junction remodels and adjusts its length to that imposed by the external load. The dashpot with viscosity $\zeta$ models dissipation with the environment. Finally, the junction is equipped with an active source of tension, which models the action of myosin motors. 

Each molecular motor produces a force of magnitude $\tilde{\beta}$. $N_\text{mot}$ motors attached to actin filaments of the junction, therefore, generate tension $T_\text{active}=\tilde{\beta}N_\text{mot}$. If the maximum possible number of attached motors is $N_\text{max}$, $T_\text{active} = \beta m$, with $\beta = \tilde{\beta}N_\text{max}$ and $m = N_\text{mot}/N_\text{max}$. The junction is, however, a part of the tissue and in the absence of perturbations the average steady state value of motors attached to all junctions $m_0N_\text{max}\neq0$. It is, therefore, appropriate to model the active tension as $T_\text{active} = \beta\left(m-m_0\right)$. This expression can also be understood as the leading-order term in the expansion of $T_\text{active}(m)$ around $m_0$ \cite{dierkes2014spontaneous}. For $m < m_0$, $T<0$, i.e. tension acts to extend the junction. While the $m_0$ term may appear counterintuitive, it reflects the fact that if motors attached to the junction are depleted, contractions of the surrounding junctions produce stronger pull on it than it can resist, resulting in elongation. 

Tension feeds back on the kinetics of association and dissociation of myosin to actin filaments, leading to the kinetic equation for $m$,
\begin{equation}
  \dot{m} = k_\text{on} - k_\text{off}\left(T\right)m, \label{eq:myo_kin}
\end{equation}
where $k_\text{on}$ is the association rate constant, assumed to be tension-independent with no limit of the total available myosin, and $k_\text{off}\left(T\right)$ is the dissociation rate constant, assumed to be a monotonously decaying function of tension $T$. The kinetics of actin-bound myosin, therefore, resembles that of a catch bond. This assumption is motivated by measurements of binding and unbinding rates of myosin motors on single actin filaments and has been shown to be described as a simple negative exponential dependence of the dissociation rate on applied tension \cite{veigel2003load,kovacs2007load}. The dynamics of the junction is given by the following set of equations,
\begin{equation}
    \zeta \dot{l} = -T + T_{\text{ext}},  \quad \tau_\text{v} \dot{l}_0 = l-l_0, \quad \tau_\text{m} \dot{m} =  1 - m F\left(T\right),  
    \label{eq:SI-one_junction} 
\end{equation}
with a natural choice being a sigmoid curve,
\begin{equation}
    F\left(T\right) = \alpha + e^{-k_0\left(T-T^*\right)}, \label{eq:F_of_t}
\end{equation}
where $\alpha > 0$ is the contribution to the myosin dissociation that does not depend on tension. The first equation describes the time evolution of the junction length due to the internal tension, $T =  k \left(l-l_0\right) + B\left(l-a\right) + \beta\left(m-m_0\right)$, and external tension, $T_\text{ext}$. $T^*$ is the threshold tension, and $k_0$ controls the steepness of the $F(T)$ curve in the vicinity of $T^*$ (fig.~\ref{fig:SI_model}C). The minus sign in front of the first term on the right-hand side indicates that $T>0$ corresponds to a junction that is contracting, i.e. $\dot{l} < 0$ for $T_\text{ext}=0$. The second equation accounts for the viscoelastic nature of the junction~\cite{clement2017viscoelastic}, i.e. the rest length $l_0$ relaxes towards the actual length $l$ with a characteristic timescale $\tau_\text{v}$. Finally, the third equation was obtained by dividing eqn.~(\ref{eq:myo_kin}) by $k_\text{on}$, where $\tau_\text{m}=1/k_\text{on}$ is the timescale of myosin association. In general, binding and unbinding of molecular motors is a stochastic process and the third equation should also include stochastic terms. For simplicity, such terms were omitted here, but were included in the two-dimensional model. Furthermore, it is assumed that $l$ and $l_0$ are comparable in magnitude, i.e. that $(l-l_0)/l_0 < 1$ which makes using a linear spring model appropriate despite the total length of the junction changing significantly as the junction collapses. 

Finally, in all simulations of the single junction model $k$, $\zeta$, and $a$ were kept fixed and, therefore, length is measured in units of $a$, time in units of $t^*=\zeta/k$, and force in units of $ka$. Parameters and their values used in the analysis of the single junction are listed in tab.~\ref{tab:single-junction}.

\subsection{Vertex Model with Active Junctions} \label{sec:SI-Vertex-Model-with-Active-Junctions}

The mechanical response of the tissue is modelled with the vertex model~\cite{farhadifar2007influence,fletcher2014vertex}. The associated mechanical energy is a function of the cell area and perimeter, 
\begin{equation}
    E_\text{VM} = \sum_\text{C}\left[\frac{\kappa_\text{C}}{2}\left(A_\text{C} - A_0\right)^2 + \frac{\Gamma_\text{C}}{2}\left(P_\text{C} - P_0\right)^2 \right], \label{eq:VM}
\end{equation}
where $A_C$ and $P_C$ are the area and the perimeter of cell $C$, respectively, $A_0$ and $P_0$ are the preferred area and perimeter, respectively (assumed to be the same for all cells) and the sum is over all cells. The first term in eqn.~(\ref{eq:VM}) accounts for three-dimensional incompressibility of cells and $\kappa_\text{C}$ is the corresponding elastic modulus of the cell $C$. The second term in eqn.~(\ref{eq:VM}) contains a combination of actomyosin contractility in the cell cortex and intercellular adhesions, where $\Gamma_\text{C}$ is the contractility modulus of cell $C$~\cite{farhadifar2007influence}. 

\begin{figure*}
    \centering
    \includegraphics[width=0.9\textwidth]{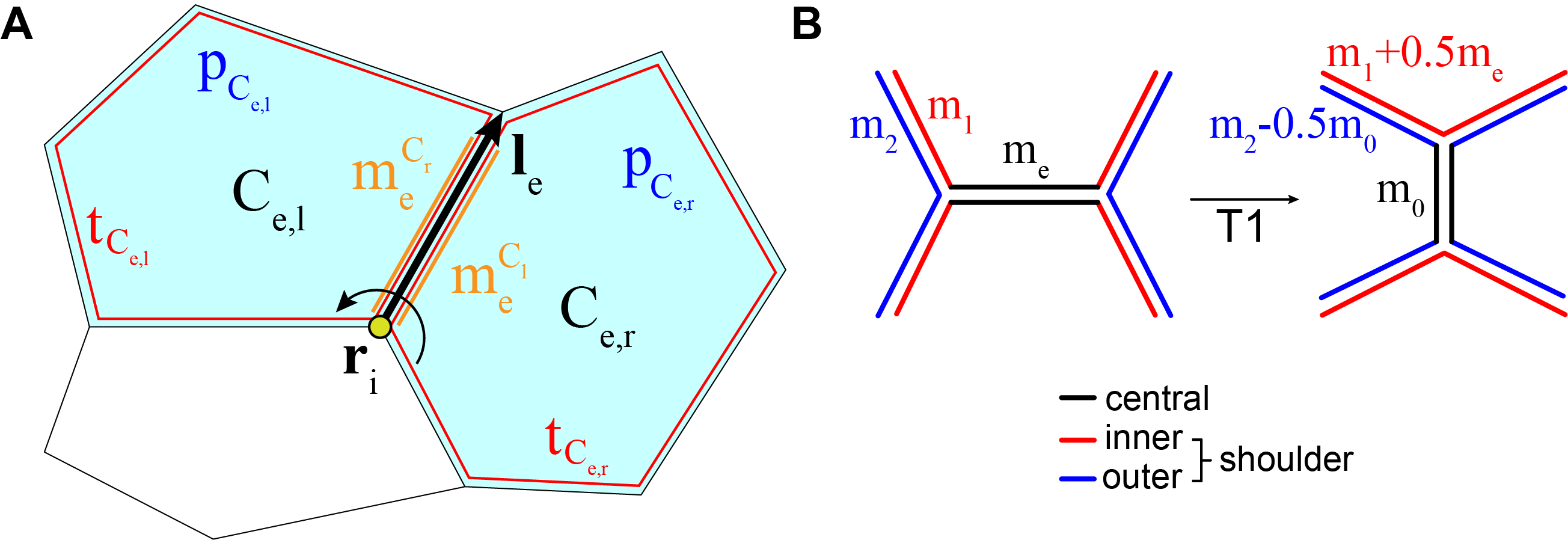}
    \caption{Schematic representation of the key ingredients in the vertex model with active junctions. \textbf{A}) The total force on vertex $i$ can be expressed as a sum of forces due to junctions, eqn.~(\ref{eq:motion_VM}). For the junction $e$, $C_\text{e,l}$ and $C_\text{e,r}$ are the cells to the left and the right, respectively, when looking in the direction of the vector $\mathbf{l}_\text{e}$. $m_\text{e}^{\text{C}_\text{l}}$ ($m_\text{e}^{\text{C}_\text{r}}$) is the myosin attached to junction $e$ due to the cell to the left (right) of it. The area ($p_{\text{C}_\text{e,l/r}}$) and perimeter ($t_{\text{C}_\text{e,l/r}}$) terms are defined in the text above. \textbf{B}) Example of the redistribution of myosin after a T1 transition for one side of the junction, and one of its inner and outer shoulder junctions. Details are given in \emph{Materials and Methods} in the main text.}
    \label{fig:AJM_schematic}
  \end{figure*}

All inertial effects were neglected and, in line with the existing literature, it is assumed that the friction can be modelled as viscous drag on each vertex. The equation of motion for vertex $i$ is, therefore, a balance between friction and mechanical forces,
\begin{equation}
    \zeta \dot{\mathbf{r}}_i = -\nabla_{\mathbf{r}_i}E_\text{VM} + \mathbf{F}_\text{active}, \label{eq:motion}
\end{equation}
where $\zeta$ is the friction coefficient, and $\mathbf{F}_\text{active}$ accounts for all active forces. Stochastic forces are, however, omitted since those do not qualitatively affect the dynamics at time scales of interest. Inserting eqn.~(\ref{eq:VM}) into eqn.~(\ref{eq:motion_VM}) leads to 
\begin{equation}
    \dot{\mathbf{r}}_i = \frac{1}{\zeta}\sum_e\left(F^\text{A}_\text{e}\mathbf{e}_\text{z}\times\mathbf{l}_\text{e}+T^\text{P}_\text{e}\hat{\mathbf{l}}_e\right)+ \frac{1}{\zeta}\mathbf{F}_\text{active}, \label{eq:motion_VM}
\end{equation}
with $F^\text{A}_\text{e}=\frac{1}{2}\left(p_{\text{C}_\text{e,l}}-p_{\text{C}_\text{e,r}}\right)$ and $T^\text{P}_\text{e}=-\left(t_{\text{C}_\text{e,l}}+t_{\text{C}_\text{e,r}}\right)$ being the magnitudes of, respectively, the area and perimeter contributions to the force due to junction $e$. The subscript $C_\text{e,l}$ ($C_\text{e,r}$) denotes the cell to the left (right) of the junction $e$ when facing in the direction of $\mathbf{l}_\text{e}$. $\mathbf{e}_\text{z} $ is the unit-length vector perpendicular to the plane of the tissue and the vector $\mathbf{l}_\text{e}$ points along the junction $e$ away from vertex $i$, $\hat{\mathbf{l}}_\text{e} = \mathbf{l}_\text{e}/l_\text{e}$ with $l_\text{e}=\left|\mathbf{l}_\text{e}\right|$, $p_{\text{C}} = -\partial E_\text{VM}/\partial A_\text{C}=-\kappa_\text{C}\left(A_\text{C} - A_0\right)$ is the hydrostatic pressure on the cell C, and $t_\text{C} = -\partial E_\text{VM}/\partial P_\text{C}=-\Gamma_\text{C}\left(P_\text{C} - P_0\right)$. Finally, the sum is over all junctions that originate at vertex $i$ and terms in the sum appear in counterclockwise order (fig.~\ref{fig:AJM_schematic}A). 

Activity is introduced by assuming that each junction contains two active elements supplied by the two cells sharing it. Furthermore, each cell is assumed to have a finite pool of myosin, $M$. Of this total myosin, a fraction $m_\text{act}^\text{C} = \sum_{e=1}^{z_\text{C}}m_\text{e}^\text{C}$ is assumed to be activated, i.e. bound the junctions, and thus depleted from the pool. Here, $z_\text{C}$ is the number of junctions shared by the cell $C$. The association rate of myosin to an individual junction is proportional to $M-m_\text{act}^\text{C}$. As in the single-junction model, the dissociation rate is proportional to the amount of myosin bound to the junction $e$, modulated by a tension-depend function $F\left(T_\text{e}\right)$. To match the steady state value of myosin of the single junction where $m_\text{eq}=F\left(T\right)^{-1}$, the prefactor of the unbinding term also needs to be $z_\text{C}$. Then, to lowest order in the contributions of the coupled junctions, $m_\text{eq}=\left(2 F\left(T\right)\right)^{-1}$. We can recover $m_\text{eq} \approx 0.5$ at $T=T^*$ by choosing a small but finite values of $\alpha$ ($\alpha=0.1$ in simulations).  This leads to the kinetic equation for the myosin motor attached to the junction $e$ by cell $C$,
\begin{equation}
    \tau_\text{m}\dot{m}_\text{e}^\text{C} = \left(M-m_\text{act}^\text{C}\right) - zm_\text{e}^\text{C}F\left(T_\text{e}\right) + \eta_\text{e}^\text{C}, \label{eq:myodyn}
\end{equation}
where it is assumed that $M$ and $z$ are same for all cells. Without loss of generality, it is possible to set $M=z$. $\tau_\text{m}$ is the inverse rate of myosin binding, i.e. the time scale of attachment of myosin motors and $\eta_\text{e}^\text{C}$ is a random white noise with zero mean and variance 
\begin{equation}
    \langle\eta_\text{e}^\text{C}\left(t\right)\eta_\text{e}^\text{C}\left(t^\prime\right)\rangle = f\delta\left(t-t^\prime\right),  \label{eq:myo_noise}  
\end{equation}
which accounts for the stochastic nature of myosin binding and unbinding. The noise term is important for the system to be able to break the symmetry imposed by using regular hexagonal tilings. It is, however, not strictly necessary to introduce noise to the myosin kinetics, but instead consider, e.g. that mechanical properties of the cells are randomly distributed. While the quantitative results would be affected, such a model is not expected to have qualitatively different behaviour compared to what is discussed here. 

Eqn.~(\ref{eq:VM}) models the passive elastic response of the tissue. In order to achieve an active T1 event, remodelling needs to be present, i.e. the system must be viscoelastic. There are various ways to include viscoelastic effects into the vertex model. In order to be consistent with the single junction model, it is further assumed that the junction $e$ has a viscoelastic contribution to the tension, $k_\text{e}\left(l_\text{e} - l_\text{e}^0\right)$, where $k_\text{e}$ is the spring constant analogue to the elastic part of the Maxwell element in the single-junction model and $l_0$ is the time-dependent rest length with dynamics,
\begin{equation}
    \tau_\text{v}\dot{l_\text{e}^0} = l_\text{e} - l_\text{e}^0, \label{eq:VM_ve}
\end{equation}
where $\tau_\text{v}$ is the characteristic time-scale for viscoelastic remodelling. The full expression for the tension of junction $e$ is,
\begin{equation}
  T_\text{e} = T^\text{P}_\text{e} + k_\text{e}\left(l_\text{e} - l_\text{e}^0\right) + \beta^{\text{C}_\text{l}}_\text{e}\left(m_\text{e}^{\text{C}_\text{l}}-m_0\right) + \beta^{\text{C}_\text{r}}_{\text{e}}\left(m^{\text{C}_\text{r}}_{\text{e}}-m_0\right). \label{eq:active_tension}
\end{equation}
As above, the superscript $C_\text{l}$ ($C_\text{r}$) denotes the cell to the left (right) of the junction $e$ when facing in the direction of $\mathbf{l}_\text{e}$. $\beta^\text{C}_\text{e}$ is the activity of the junction $e$ produced by the cell $C$, i.e. it is a constant with units of force that measures the strength of the mechanochemical coupling and $m_0$ has the same meaning as in the single-junction model. For simplicity, $\Gamma_\text{C}\equiv\Gamma$ for all cells and $k_\text{e}\equiv k$ for all junctions and both parameters were kept constant in all simulations. The unit of length, $a$, was chosen to be the length of the side of a regular hexagon, which allows us to set the unit of time $t^*=\zeta/\left(\Gamma + k\right)$, and the unit of force $f^*=\left(\Gamma + k\right)a$. Eqn.~(\ref{eq:motion_VM}), with $T_\text{e}^\text{P}$ given by eqn.~(\ref{eq:active_tension}), and eqns.~(\ref{eq:myodyn}) and (\ref{eq:VM_ve}) describe the dynamics of the vertex model with active junctions.

Parameters and their values used in the analysis of the vertex model with active junctions are listed in tab.~\ref{tab:ajm}.

\subsection{Characterisation of T1 transitions and tissue flow}
\label{sec:Characterisation-of-the-T1-transition}

To identify and characterise T1 transitions, and quantify the associated deformation of the model tissue, we used the analysis method introduced by Graner, et al.\ \cite{graner2008discrete}. For cellular patterns, three tensors are defined, texture ($\hat{\boldsymbol{M}}$), geometrical texture change ($\hat{\boldsymbol{B}}$), and topological texture change ($\hat{\boldsymbol{T}}$). Tensor $\hat{\boldsymbol{M}}$ describes the shape of the current cell configuration and it is defined as
\begin{equation}
    \hat{\boldsymbol{M}} = \langle\hat{\boldsymbol{m}}\rangle = \langle\boldsymbol{\ell}\otimes\boldsymbol{\ell}\rangle=
    \begin{pmatrix}
        \langle X^2 \rangle & \langle XY \rangle \\
        \langle YX  \rangle & \langle Y^2 \rangle 
    \end{pmatrix}, \label{eq:Mtensor}
\end{equation}
where $X$ ($Y$) is the $x$ ($y$) component of the vector $\boldsymbol{\ell} = \mathbf{r}_2 - \mathbf{r}_1$ connecting centroids of two neighbouring cells at positions $\mathbf{r}_1$ and $\mathbf{r}_2$, respectively, $\langle\cdot\rangle=\frac{1}{N_\text{tot}}\sum\left(\cdot\right)$ is an average over $N_\text{tot}$ pairs of neighbours, i.e. cell-cell contacts, and $\hat{\boldsymbol{m}} = \boldsymbol{\ell}\otimes\boldsymbol{\ell}$. Using the same notation, the tensor $\hat{\boldsymbol{B}}$ describes shape changes of the cell configuration during a time interval $\Delta t$ and it is defined as
\begin{equation}
    \hat{\boldsymbol{B}} = \hat{\boldsymbol{C}} + \hat{\boldsymbol{C}}^\text{T},\label{eq:Ctensor}
\end{equation}
where the superscript T denotes the matrix transpose and
\begin{equation}
    \hat{\boldsymbol{C}} = \left\langle\boldsymbol{\ell}\otimes\frac{\Delta\boldsymbol{\ell}}{\Delta t}\right\rangle =  
    \begin{pmatrix}
        \left\langle X\frac{\Delta X}{\Delta t}\right\rangle  & \left\langle Y\frac{\Delta X}{\Delta t}\right\rangle  \\
        \left\langle X\frac{\Delta Y}{\Delta t}\right\rangle  & \left\langle Y\frac{\Delta Y}{\Delta t}\right\rangle  
    \end{pmatrix}. \label{eq:Btensor}
\end{equation}
Finally, the tensor $\hat{\boldsymbol{T}}$ identifies T1 transitions by quantifying topological changes of the cell configuration in the time interval $\Delta t$ via tracking appearance and disappearance of contacts between cells. It is defined as
\begin{equation}
    \hat{\boldsymbol{T}} = \frac{1}{\Delta t}\langle \hat{\boldsymbol{m}}\rangle_\text{a} - \frac{1}{\Delta t}\langle \hat{\boldsymbol{m}}\rangle_\text{d}, \label{eq:Ttensor}
\end{equation} 
where $\langle\cdot\rangle_\text{a}$ ($\langle\cdot\rangle_\text{d}$) is the average over contacts that appeared (disappeared) during the time interval $\Delta t$. The total number of contacts that appear and disappear is typically much smaller than $N_\text{tot}$, which means that $\hat{\boldsymbol{T}}$ data can be quite noisy. While all three tensors can be calculated for individual cells, the averaging is meant to be carried over a mesoscopic region. We, therefore, averaged over the 14 central active cells (dark-shaded cells in fig.\ \ref{fig:T1_schematic}A in the main text), as well as over an ensemble of $N=32$ noise realisations. For the disordered tissue, we averaged over all $N=520$ active cells.

\begin{figure*}
\centering
\includegraphics[width=1.0\textwidth]{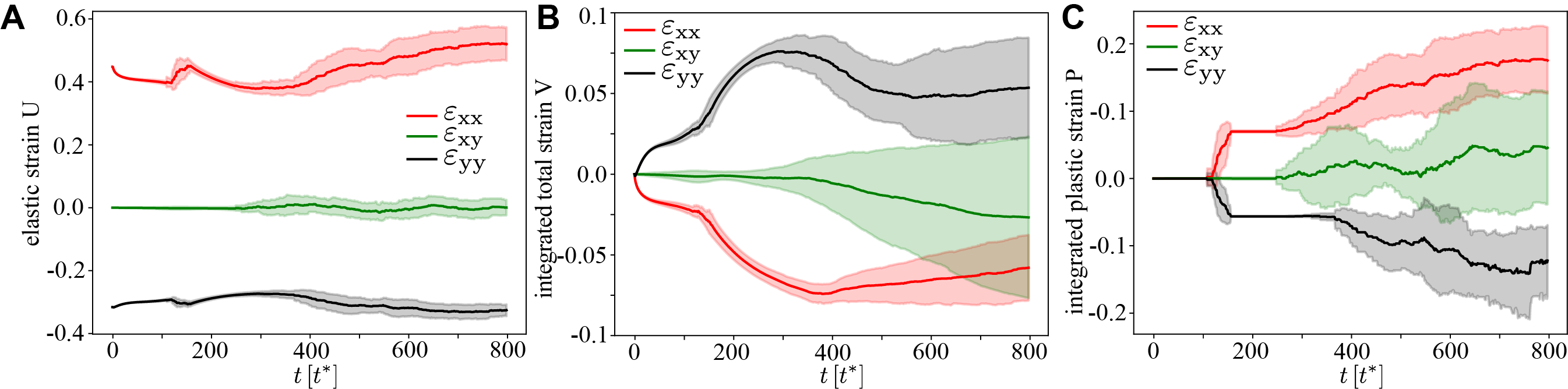}
\caption{Continuous strain tensors through the active T1 transition, for $\beta=0.8f^*$, $f_{\text{pull}} = 0.15 f^*$, and the other parameters corresponding to figs.~\ref{fig:T1_schematic} and \ref{fig:T1_dynamics} in the main text. {\bf{A}}) Statistical strain tensor $ \hat{\boldsymbol{U}}$ defined in eqn.~(\ref{eq:Utensor}), with the initial undeformed state used as reference configuration. {\bf{B}}) Total strain, i.e. integrated $ \hat{\boldsymbol{V}}$ tensor defined in eqn.~(\ref{eq:tot_strain}).  {\bf{C}}) Integrated plastic strain tensor $\hat{\boldsymbol{P}}$ defined in eqn.~(\ref{eq:Ptensor}). One can see that the active T1 proceeds at near constant elastic strain with cells elongated in the pulling direction, i.e. $\varepsilon_\text{xx}>0$ and $\varepsilon_\text{yy}<0$. Conversely, there is clear convergence-extension since $\varepsilon_\text{xx}<0$ and $\varepsilon_\text{yy}>0$ in the total strain.  Here the traces are an average over $n=32$ simulations, and the shading corresponds to the error of the mean.}
\label{fig:Graner_traces}
\end{figure*}

Tensors $\hat{\boldsymbol{M}}$, $\hat{\boldsymbol{B}}$, and $\hat{\boldsymbol{T}}$ all involve averaging over cell-cell contacts and, therefore, describe a discrete system. In order to make connections to continuous deformations of the entire tissue, one introduces their continuos counterparts, the statistical strain tensor ($\hat{\boldsymbol{U}})$, the velocity gradient tensor ($\hat{\boldsymbol{V}})$, and the tensor of the rate plastic deformations (i.e. topological rearrangements rate) ($\hat{\boldsymbol{P}}$) \cite{graner2008discrete}. These tensors are defined as
\begin{equation}
    \hat{\boldsymbol{U}} = \frac{1}{2}\left(\log\hat{\boldsymbol{M}}-\log\hat{\boldsymbol{M}_0}\right), \label{eq:Utensor}
\end{equation}  
where $\hat{\boldsymbol{M}_0}$ is the texture tensor of an arbitrary reference configuration for the statistical relative strain - we chose the initial undeformed configuration. Further,
\begin{equation}
    \hat{\boldsymbol{V}} = \frac{1}{2}\left(\hat{\boldsymbol{M}}^{-1}\hat{\boldsymbol{C}}+\hat{\boldsymbol{C}}^{\text{T}}\hat{\boldsymbol{M}}^{-1}\right), \label{eq:Vtensor}
\end{equation}
and
\begin{equation}
    \hat{\boldsymbol{P}} = \frac{1}{2}\left(\hat{\boldsymbol{M}}^{-1}\hat{\boldsymbol{T}}+\hat{\boldsymbol{T}}\hat{\boldsymbol{M}}^{-1}\right). \label{eq:Ptensor}
\end{equation}
In the case when variations in $N_\text{tot}$ can be neglected, one can show that \cite{graner2008discrete},
\begin{equation}
    \hat{\boldsymbol{V}} = \frac{\mathcal{D}\hat{\boldsymbol{U}}}{\mathcal{D}t} + \hat{\boldsymbol{P}}, \label{eq:V_U_P}
\end{equation}
where $\mathcal{D}/\mathcal{D}t$ is the corotational derivative \cite{larson1999structure}. This equation just states that the velocity gradient is a sum of two contributions, reversible changes of the internal strain and the rate of irreversible plastic rearrangements. The significance of eqn.~(\ref{eq:V_U_P}) is that if one assumes that there are no plastic events other than T1 transitions, it is possible to obtain the total strain of the system $\hat{\boldsymbol{\varepsilon}}^\text{tot}$ by integrating the tensor $\hat{\boldsymbol{V}}$ over time, i.e.
\begin{equation}
    \hat{\boldsymbol{\varepsilon}}^\text{tot} = \int_{t_0}^{t_\text{tot}} \mathrm{d}t^\prime \hat{\boldsymbol{V}}\left(t^\prime\right), \label{eq:tot_strain}
\end{equation}
where $t_0$ is the time when the activity is switched on and $t_\text{tot}$ is the total simulation time. The difference $\varepsilon^\text{tot}_\text{xx} - \varepsilon^\text{tot}_\text{yy}$ of the $xx$ and $yy$ components of $\hat{\boldsymbol{\varepsilon}}^\text{tot}$ (i.e. total strains in the $x$ and $y$ direction, respectively) was used as the measure the amount of convergence-extension in the tissue induced by the active T1 transition. Figure \ref{fig:Graner_traces} shows ensemble-averaged time traces of $\hat{\boldsymbol{U}}$ and the time integrated total strain $\hat{\boldsymbol{V}}$ and plastic strain $\hat{\boldsymbol{P}}$ through the active T1 transition and the subsequent propagation phase.

\begin{figure*}
\centering
\includegraphics[width=1.0\textwidth]{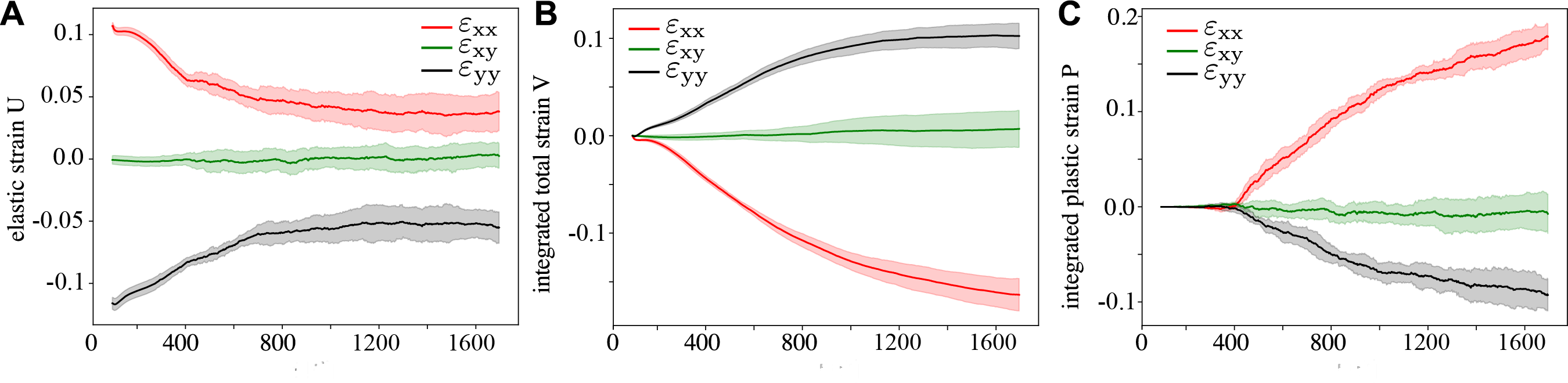}
\caption{Continuous strain tensors for the fully active tissue at $\beta=0.5f^*$, $f_{\text{pull}} = 0.2 f^*$, and the other parameters corresponding to figs.~\ref{fig:T1_schematic} and \ref{fig:T1_dynamics} in the main text. {\bf{A}}) Statistical strain tensor $ \hat{\boldsymbol{U}}$ defined in eqn.~(\ref{eq:Utensor}), with the initial undeformed state used as reference configuration. {\bf{B}}) Total strain, i.e. integrated $ \hat{\boldsymbol{V}}$ tensor defined in eqn.~(\ref{eq:tot_strain}).  {\bf{C}}) Integrated plastic strain tensor $\hat{\boldsymbol{P}}$ defined in eqn.~(\ref{eq:Ptensor}). Convergence-extension flow proceeds with cells elongated in the pulling direction, i.e. $\varepsilon_\text{xx}>0$ and $\varepsilon_\text{yy}<0$ while $\varepsilon_\text{xx}<0$ and $\varepsilon_\text{yy}>0$ in the total strain.  Here the traces are an average over $n=5$ simulations, and the shading corresponds to the standard deviation.}
\label{fig:Graner_traces_patch}
\end{figure*}

\subsection{Convergence-extension in a patch of randomly shaped active cells}
\label{Sec:SI-CE-random}
We used the two-dimensional model with parameters for the cell mechanics and the myosin feedback loop as defined in tab.~\ref{tab:ajm} on a tissue patch with cells of random shapes. The patch was generated using an iterative procedure. First, $N=600$ points were placed at random in a square region of size $L$ chosen such that the average cell area is equal to that of the hexagonal patch, i.e. $A_0 = 2.598a^2$. During the initialisation process, we ensured that no two points are closer than $a$ to each other. Once all points are placed in the box, the Voronoi diagram was constructed and centroids of each cell of the Voronoi diagrams were used as the seeds for constructing Voronoi diagram in the next iteration. This was repeated until no centroid moved more than $5\times10^{-5}a$ between two consecutive iterations resulting in a centroidal Voronoi tessellation. Centroidal Voronoi tessellations have several convenient properties. For example, typically, there are no outliers (i.e. very large or very small cells are unlikely) and the distribution of cell neighbours is remarkably similar to actual epithelia. Finally, to avoid the spontaneous formation of an actomyosin cable at the outside border, we used a setup where the $N=520$ interior cells are active, and a one-cell thick layer of passive cells forms the boundary.  

We used the same stretching protocol as for the ordered patch. At every vertex on the left and right boundaries, the force $f_{\text{pull}}$ was exerted during the entire duration of the simulation. The tissue was first passively polarised by stretching it for $10^3t^*$ with activity and the viscoelastic relaxation turned off. We then turned on activity and viscoelasticity and simulated the system for $1.6\times10^3t^*$. This protocol was repeated for a range of activities $\beta$, viscous relaxation times $\tau_\text{v}$ and myosin times $\tau_\text{m}$ (not shown since like we observed for the single active T1, results for convergence-extension were largely independent of $\tau_\text{v}$ and $\tau_\text{m}$ and quantitatively similar to fig.~\ref{fig:random_patches} in the main text). All results were averaged over $n=5-33$ different random initial configurations and with different random number generator seeds for the myosin noise. The results shown in fig.~\ref{fig:random_patches}C-D of the main text are reported with a 95\% confidence interval computed using bootstrapping by resampling $10^3$ times.

Figure \ref{fig:random_patches} of the main text shows tissue dynamics for $\beta=0.5f^*$ and $f_\text{pull}=0.2f^*$, which is the optimal region for convergence-extension. We measured convergence-extension using the integral of $\hat{\mathbf{V}}$, where the calculation was done over the entire active region, and the starting point was the time when the activity was turned on. For the elastic strain $\hat{\mathbf{U}}$, we used the undeformed disordered initial condition to define the reference strain $\hat{\mathbf{M}}_0$. Figure \ref{fig:Graner_traces_patch}, analogous to fig.~\ref{fig:Graner_traces} for the single T1 event, shows the time traces of the ensemble averaged tensors for elastic strain $\hat{\mathbf{U}}$, integrated total strain $\hat{\mathbf{V}}$, and integrated plastic strain $\hat{\mathbf{P}}$.

We also quantified the amount of tension and myosin polarisation using the tissue-averaged stress tensor components 
\begin{equation}
\hat{\mathbf{T}}_\text{tension}=\large\langle\frac{1}{A_\text{C}}\sum_{\text{e}\in\text{C}}T_\text{e}\hat{\mathbf{l}}_\text{e}\times\mathbf{l}_\text{e}\large\rangle \label{eq:SI-T_tension}
\end{equation}
and 
\begin{equation}
    \hat{\mathbf{M}}_\text{myo}=\large\langle\frac{1}{A_\text{C}}\sum_{\text{e}\in\text{C}}m_\text{e}\hat{\mathbf{l}}_\text{e}\times\mathbf{l}_\text{e}\large\rangle.   \label{eq:SI-M_myo}  
\end{equation}

To measure convergence-extension strain and shape, myosin and tension polarisations, we chose $t = 700t^*$, which is the time point where convergence-extension and polarisation stabilise, giving the results shown in figs.~\ref{fig:random_patches}C,D of the main text.

\begin{figure*}
\centering
\includegraphics[width=1.0\textwidth]{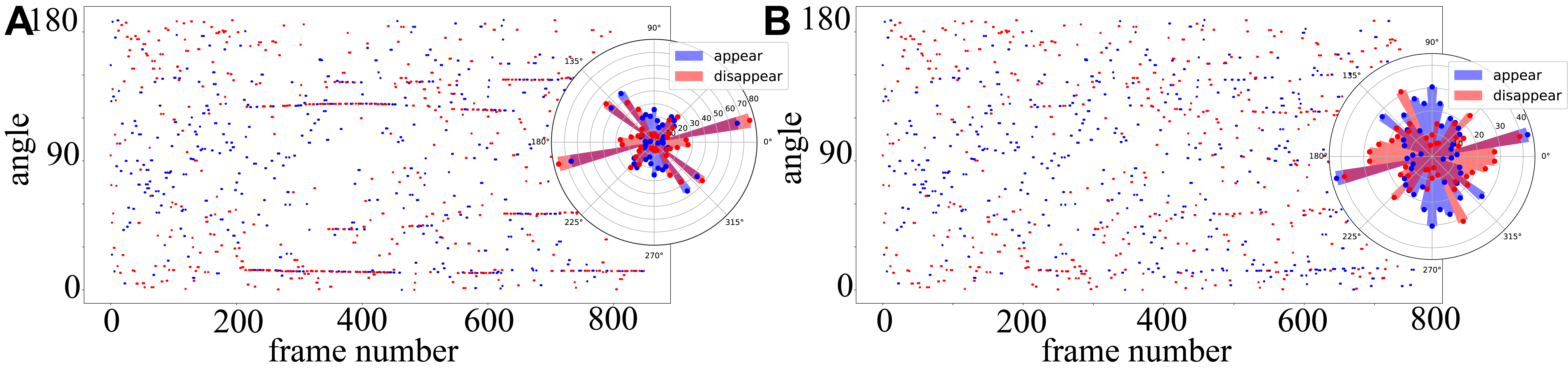}
\caption{Measuring T1 transitions in the fully active random patch. A representative example taken from a sample of $n=5$ simulations for $\beta=0.5$, $f_{\text{pull}}=0.2$. {\bf A}) Appearing and disappearing junctions as a function of simulation frame number and angle, before filtering. Inset: Raw T1 histogram. {\bf B}) Same dataset after filtering procedure.}
\label{fig:T1filter}
\end{figure*}

To measure orientations of T1s, we first identified cells involved in a T1 transition by diagonalising the $\hat{\mathbf{T}}$ tensor associated with active cells at a given instance in time. The signature of a T1 event is a non-zero $\hat{\mathbf{T}}$, and the sign of its trace determines if a junction appeared or disappeared. The eigenvector corresponding the largest eigenvalue determines the direction of the event. We tracked all of such events over the simulation runtime and generated polar histograms such as the one shown in fig.~\ref{fig:random_patches}B of the main text. As shown in fig.~\ref{fig:T1filter} (left) for $\beta=0.5f^*$ and $f_{pull}=0.2f^*$, there are temporally strongly correlated back-and-forth T1 transitions at the same angle. These correspond to four cells flipping back and forth through a T1 transition, or ``flickering'' in the movies. As these events are artefacts of the simulation, we excluded them from the data. Figure \ref{fig:T1filter}B, which was averaged over $n=5$ independent simulations, shows that flickering events have mostly been filtered out of the dataset. 

\begin{figure*}
\centering
\includegraphics[width=1.0\textwidth]{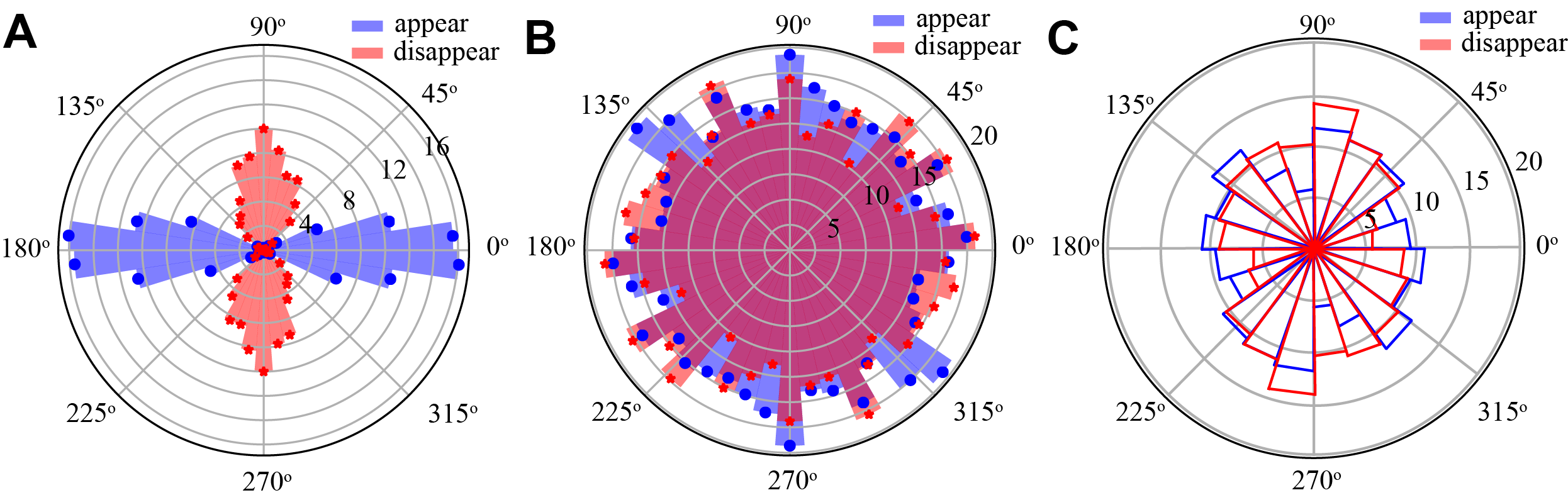}
\caption{T1 histograms in the parameter range where no convergence-extension occurs. {\bf A}) Passive tissue at $\beta=0$ pulled with $f_{\text{pull}}=0.2$, T1s are passive and aligned along pulling direction. {\bf B}) Active tissue at $\beta=0.5$ with no applied force $f_{\text{pull}}=0.0$, here T1s are distributed isotropically. {\bf C}) Experimental T1 distribution in the anterior region of the streak, strongly resembling the active isotropic case.}
\label{fig:T1other}
\end{figure*}

Figure \ref{fig:T1other} shows filtered T1 histograms for other mechanical conditions, outside the parameter region of where convergence-extension occurs, averaged over $n=5$ independent simulations. In particular, we also include a passive tissue that flows in the direction of the pulling, and an active isotropic tissue without applied forces where the T1 distribution is isotropic. This last situation strongly resembles the observed T1 distribution in the anterior region of the streak (see fig.~\ref{fig:T1other}C). As shown in fig.~\ref{fig:UV_anterior}, this is also a mostly isotropic tissue.

\clearpage
\newpage

\section{Experimental data analysis} \label{sec:SI-experiment-data-analysis}

\subsection{Experimental data}

Active myosin (Phosphorylated myosin light chain) and actin staining in fixed embryos was performed as described in \cite{rozbicki2015myosin}. Embryos of a transgenic chick line with cell membranes of all cells in the embryonic and extra embryonic tissues labelled with a green fluorescent protein tag (myr-EGFP) were live-imaged using a dedicated light-sheet microscope as described previously \cite{rozbicki2015myosin}. The microscope produces cell-resolution images of the entire embryo with time resolution of 3 min between frames for periods up to 16 hours (stage EGXIII-HH4). $n=6$ rectangular areas of interest of size $\approx255\times220$ $\mu m^2$ were chosen to lie next to each other along the sickle-shaped mesendoderm precursor region in the embryo's posterior, perpendicular to the direction of the forming primitive streak. The two central sections were chosen to lie in the middle of the sickle region that initiates the formation of the streak. Each area of interest was tracked for $\approx 4.6$ h covering the onset of the flows driving streak formation. The average motion of the area of interest, determined via particle image velocimetry using PIVlab \cite{thielicke2021particle}, was used to track its displacement to be able to follow the same patch of cells over time. After this, these image time series were bandpass filtered to remove some noise followed by segmentation using the watershed algorithm in MATLAB. To follow the same cells over time, their centroid positions calculated form the segmentation of the first image of the time series are projected forward to the next frame using a newly calculated high resolution velocity field between these successive images and using these as seed points for the segmentation of the next image \cite{rozbicki2015myosin}. This procedure allowed us to track individual cells between consecutive time frames and determine changes in neighbours over time as well as determine the directions of appearing and disappearing junctions associated with T1 transitions. The  $\hat{\mathbf{M}}$, $\hat{\mathbf{U}}$, and $\hat{\mathbf{V}}$ tensors in a particular area of interest were averaged for all cells in centred circular domains  of  $190$ $\mu$m diameter between successive pairs of segmented images and averaged over 30 minute time intervals.

The spatially averaged $\hat{\mathbf{M}}(t)$ texture tensor allowed us to both compute $\hat{\mathbf{U}}$ and also to directly quantify shape polarisation from the eigenvalues $\left(m_L,m_S\right)$ and eigenvectors  $\left(\xi_L^M,\xi_S^M\right)$ of the time-averaged $\langle \hat{\mathbf{M}}(t)\rangle_t$, where $L$ and $S$ label the large and small components, respectively. We defined the dimensionless shape polarisation shown in fig.~\ref{fig:experiment}C in the main text as 
\begin{equation} 
    p_S = \frac{m_L-m_S}{m_L+m_s}, 
\end{equation}
and the shape polarisation direction as the angle $\xi_L^M$ makes with the lab frame $x$ axis.

\begin{figure*}
    \includegraphics[width=0.95\textwidth]{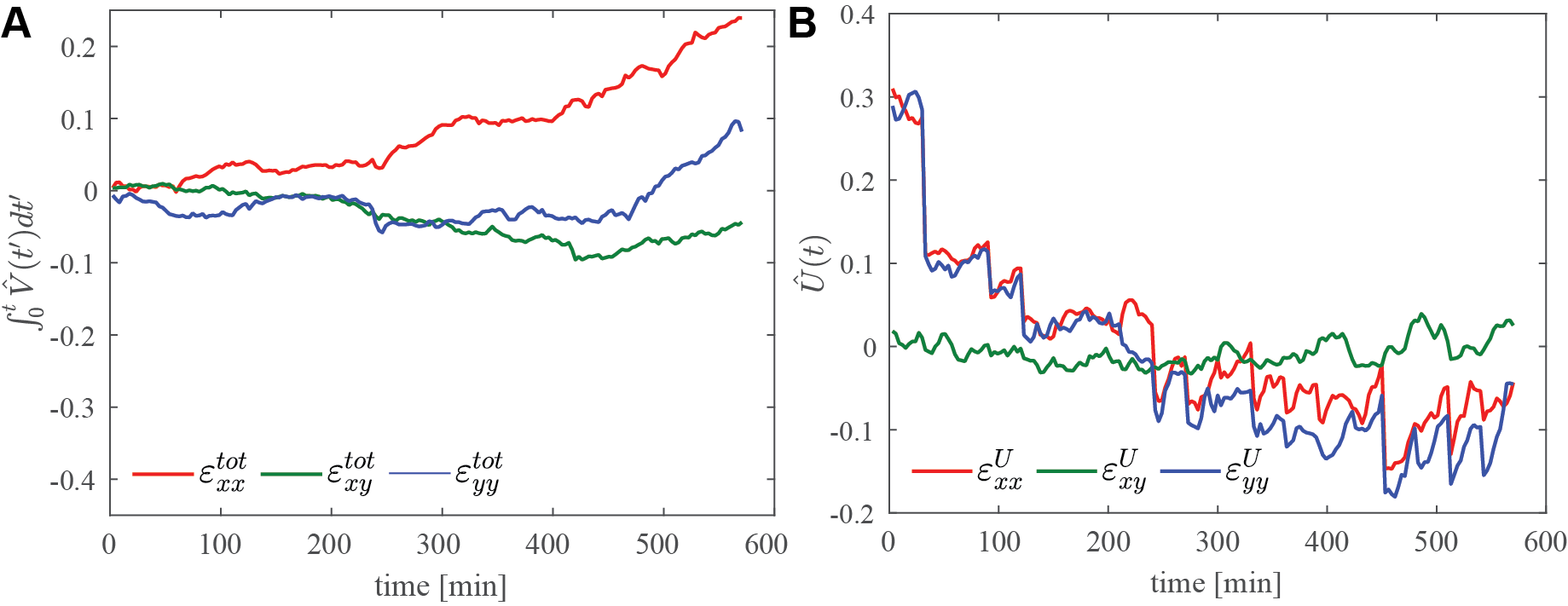}
\caption{Region of interest in the anterior of the embryo. {\bf A}) Integrated total strain $\hat{\boldsymbol{\varepsilon}}^\text{tot}(t)$ and {\bf B}) elastic strain $\hat{\mathbf{U}}(t)$. }
\label{fig:UV_anterior}
\end{figure*}

Unlike in the simulation, in the real embryo, cells divide, ingress, and flow in and out of the region of interest. Therefore, it is not immediately clear what to use as the reference texture tensor $\hat{\mathbf{M}}_0$ when computing $\hat{\mathbf{U}}$ tensor. For simplicity, we chose the isotropic tensor constructed from the time-averaged eigenvalues as 
\begin{equation} 
    \hat{\mathbf{M}}_0 = \frac{1}{2}(m_s+m_L) \hat{\mathbf{I}}. 
\end{equation}
To compute the flow polarisation, we measured the integrated total strain tensor
\begin{equation}
    \hat{\boldsymbol{\varepsilon}}^\text{tot}(t) = \int_{t_0}^{t} \mathrm{d}t^\prime \hat{\boldsymbol{V}}\left(t^\prime\right), 
\end{equation}
analogous to the simulated tissue patches. Similar to $\hat{\mathbf{M}}$, we can then compute the eigenvalues $(V_L,V_S)$ and eigenvectors  $(\xi_L^V,\xi_S^V)$ of the final $\hat{\boldsymbol{\varepsilon}}^\text{tot}(t_{max}) $, where we again label the large and small components $L$ and $S$, respectively. Here we now typically have $V_S<0$ and $V_L>0$, corresponding to the convergence and extension directions of the tissue, respectively. We compute the total flow magnitude in fig.~\ref{fig:experiment}C in the main text as
\begin{equation} 
    \varepsilon^\text{tot}_{C-E} = V_L-V_S , 
\end{equation}
and the flow polarisation direction as the angle $\xi_L^V$ makes with the lab frame $x-$axis.
For comparison, in fig.~\ref{fig:UV_anterior} we show the integrated $\hat{\boldsymbol{V}}$ and the $\hat{\mathbf{U}}$ tensor for a region in the anterior of the embryo which does not undergo convergence-extension.

\clearpage

\begin{table}[tb]
  \caption{Values of the parameters in the single junction model. Units: length ($a$), time ($t^*=\zeta/k$), force ($ka$). }
  \centering
\begin{tabular}{c c c}
 \multicolumn{3}{c}{Base} \\
 \hline\hline
   Parameter & \multicolumn{2}{c}{Description}  \\ [0.5ex] 
\hline
   $k$             & \multicolumn{2}{c}{spring constant} \\
   $a$             & \multicolumn{2}{c}{barrier rest length} \\
   $\zeta$         & \multicolumn{2}{c}{friction with substrate}\\ [1.5ex]
\multicolumn{3}{c}{Model}\\
\hline\hline
   Parameter & Description  & Value range\\ [0.5ex]
\hline
   $B$             & barrier spring constant & $0-0.2k$\\
   $T_\text{ext}$  & applied external tension  & $0-1ka$ \\
   $\beta$         & myosin activity & $0-3ka$ \\
   $\tau_\text{v}$ & viscoelastic time  & $10t^*$\\ 
   $\tau_\text{m}$ & myosin time & $10t^*$\\
   $m_0$           & myosin reference level & 0.5 \\
   $T^*$           & threshold tension &  $0.3ka$\\
   $k_0$           & slope of $m$ vs.~$T$ at $T^*$ & $2/T^*$ \\
   $\alpha$        & tension-independent myosin dissociation & 1 \\
   %\bottomrule
\end{tabular}
\label{tab:single-junction}
\end{table}

\begin{table}[htb]
  \caption{Values of parameters used in the vertex model with active junctions. Units: length ($a$), time ($t^*=\zeta/\left(\Gamma+k\right)$), force ($f^*=\left(\Gamma+k\right)a$).}
  \centering
\begin{tabular}{c c c}
 \multicolumn{3}{c}{Base} \\
 \hline\hline
   Parameter & \multicolumn{2}{c}{Description}  \\ [0.5ex] 
\hline
   $a$             & \multicolumn{2}{c}{hexagonal cell edge length} \\
   $\Gamma$        & \multicolumn{2}{c}{perimeter modulus} \\
   $k$             & \multicolumn{2}{c}{spring constant} \\
   $\zeta$         & \multicolumn{2}{c}{friction with substrate}\\ [1.5ex]
\multicolumn{3}{c}{Model}\\
\hline\hline
   Parameter & Description  & Value range\\ [0.5ex]
\hline
   $\kappa$        & area modulus &  $1f^*/a^3$ \\
   $A_0$           & target cell area  & $3\sqrt{3}a^2/2$ \\
   $P_0$           & target cell perimeter &  $6a$ \\
   $f_\text{pull}$ & pulling force  & $0.0-0.3f^*$ \\
   $\beta$         & myosin activity & $0.0-1.4f^*$ \\
   $\tau_\text{v}$ & viscoelastic time  & $10^0-10^3t^*$\\ 
   $\tau_\text{m}$ & myosin time & $10^1-10^3t^*$\\
   $T^*$           & threshold tension &  $0.3f^*$\\
   $k_0$           & slope of $m$ vs.~$T$ at $T^*$ & $2/T^*$ \\
   $m_0$           & myosin reference level & 0.5 \\
   $M$             & total cell myosin & 6\\ 
   $f$             & variance of myosin fluctuations & $1$ \\
   $\alpha$        & tension-independent myosin dissociation & 0.1 \\
   %\bottomrule
\end{tabular}
\label{tab:ajm} 
\end{table}

\section*{Movies}
\section*{Effects of activity vs. external pulling for fixed $\tau_\text{m}$ and $\tau_\text{v}$.}

S1\_movie\_beta\_0.0\_fpull\_0.15.mp4; passive system being pulled with $f_\text{pull}=0.15f^*$.

S2\_movie\_beta\_0.4\_fpull\_0.15.mp4; active system with $\beta=0.4f^*$ for the four central cells being pulled with $f_\text{pull}=0.15f^*$, $\tau_\text{m}=100t^*$, and $\tau_\text{v}=20t^*$.

S3\_movie\_beta\_0.8\_fpul\_0.15.mp4; active system with $\beta=0.8f^*$ for the four central cells being pulled with $f_\text{pull}=0.15f^*$, $\tau_\text{m}=100t^*$, and $\tau_\text{v}=20t^*$.

S4\_movie\_beta\_0.8\_fpull\_0.05.mp4; active system with $\beta=0.8f^*$ for the four central cells being pulled with $f_\text{pull}=0.05f^*$, $\tau_\text{m}=100t^*$, and $\tau_\text{v}=20t^*$.

S5\_movie\_beta\_0.8\_fpull\_0.25.mp4; active system with $\beta=0.8f^*$ for the four central cells being pulled with $f_\text{pull}=0.25f^*$, $\tau_\text{m}=100t^*$, and $\tau_\text{v}=20t^*$.

S6\_movie\_beta\_1.2\_fpull\_0.15.mp4, active system with $\beta=1.2f^*$ for the four central cells being pulled with $f_\text{pull}=0.15f^*$, $\tau_\text{m}=100t^*$, and $\tau_\text{v}=20t^*$.

\section*{Effects of $\tau_\text{m}$ and $\tau_\text{v}$ for fixed activity $\beta$ and external pulling $f_\text{pull}$.}

S7\_movie\_taum\_20\_tauv\_20.mp4, active system with $\beta=1.0f^*$ and external pulling force $f_\text{pull}=0.15f^*$ with $\tau_\text{m}=20t^*$ and $\tau_\text{v}=20t^*$.

S8\_movie\_taum\_20\_tauv\_100.mp4, active system with $\beta=1.0f^*$ and external pulling force $f_\text{pull}=0.15f^*$ with $\tau_\text{m}=20t^*$ and $\tau_\text{v}=100t^*$.

S9\_movie\_taum\_100\_tauv\_20.mp4, active system with $\beta=1.0f^*$ and external pulling force $f_\text{pull}=0.15f^*$ with $\tau_\text{m}=100t^*$ and $\tau_\text{v}=20t^*$.

S10\_movie\_taum\_100\_tauv\_100.mp4, active system with $\beta=1.0f^*$ and external pulling force $f_\text{pull}=0.15f^*$ with $\tau_\text{m}=100t^*$ and $\tau_\text{v}=100t^*$.

S11\_movie\_taum\_100\_tauv\_500.mp4, active system with $\beta=1.0f^*$ and external pulling force $f_\text{pull}=0.15f^*$ with $\tau_\text{m}=100t^*$ and $\tau_\text{v}=500t^*$.

S12\_movie\_taum\_500\_tauv\_100.mp4, active system with $\beta=1.0f^*$ and external pulling force $f_\text{pull}=0.15f^*$ with $\tau_\text{m}=500t^*$ and $\tau_\text{v}=100t^*$.

S13\_movie\_taum\_500\_tauv\_500.mp4, active system with $\beta=1.0f^*$ and external pulling force $f_\text{pull}=0.15f^*$ with $\tau_\text{m}=500t^*$ and $\tau_\text{v}=500t^*$.

\section*{Random patch}

S14\_random\_fpull\_0.2\_beta\_0.5.mp4, patch with random active cells with $\beta=0.5f^*$ and external pulling force $f_\text{pull}=0.2f^*$.

\section*{Live images of chick embryo}
S15\_movie\_base\_of\_streak.avi, image sequence taken at base of the forming streak showing cell intercalations. Centres of some cells are labelled with coloured dots for easier identification. Crossed blue and red lines indicate individual intercalation events. Blue lines indicate direction of junction contraction and red lines direction of extension of newly formed junctions. White scale bar 50 $\mu$m.

S16\_movie\_anterior\_of\_streak.avi, image sequence taken in region of epiblast in front of forming streak. Centres of some cells are labelled with coloured dots for easier identification. Crossed blue and red lines indicate individual intercalation events. Blue lines indicate direction of junction contraction and red lines direction of extension of newly formed junctions. Images are taken at 3 minute intervals.White scale bar 50 $\mu$m.

\end{document}